\documentclass[]{article}
\usepackage{graphicx}
\usepackage{cite}
\usepackage{subfigure}
\usepackage{amsmath}
\usepackage{amssymb}
\usepackage{fancyhdr}
\usepackage{url}
\usepackage{algorithm}
\usepackage{algorithmic}
\usepackage{xspace}
\usepackage{amsthm}

\usepackage{color}
\usepackage{layouts}  %helps print textheight and textwidth in inches...it's about 8.5 but I'm only getting 6 in height inside a figure environment
\usepackage{multirow}
\usepackage{rotating}  % To put figures with their captions sideways
\usepackage[top=2.0in, bottom=1.5in, left=1in, right=1in]{geometry}

\usepackage[bookmarks=false,colorlinks=true,linkcolor={blue},citecolor={blue},urlcolor={blue}]{hyperref}

\theoremstyle{definition}
\newtheorem{exmp}{Example}[section]

\newcommand{\moocDB}{MOOCdb\xspace}

\newcommand{\pii}{personally identifiable information}

\begin{document}
\begin{titlepage}
\begin{center}
%{ \large \bfseries Methods and Tools for Analysis of Data from MOOCs:}
{ \huge MOOCdb: \\ Developing Standards and Systems to support MOOC Data Science}
\vspace{4cm}
\end{center}
\end{titlepage}

\section*{Authors}
{\Large  Kalyan Veeramachaneni $^1$}
\vspace{0.25cm}\\
{\Large Sherif Halawa $^3$} 
\vspace{0.25cm}\\
{\Large Franck Dernoncourt $^1$} 
\vspace{0.25cm}\\
{\Large Una-May O'Reilly$^1$}
\vspace{0.25cm}\\
{\Large Colin Taylor $^1$}
\vspace{0.25cm}\\
{\Large Chuong B. Do$^2$}

\vspace{2cm}
\noindent {\Large Computer Science and Artificial Intelligence Laboratory}
\vspace{0.12cm}\\
{\LARGE Massachusetts Institute of Technology}
\vspace{1cm}\\

\noindent{\LARGE Coursera, Inc.} 
\vspace{1cm}\\

\noindent {\Large Department of Electrical Engineering} 
\vspace{0.12cm}\\
{\LARGE Stanford University} \\

\section*{Acknowledgements}
Authors would like to acknowledge of James Tauber (edX), Carlos Andres Rocha (edX), Zach Pardos (University of California Berkeley), Jennifer DeBoer (TLL), Sebastian Leon (MIT), Piotr Mitros (edX)  for their contributions in refining the schema. This work has also been supported by a number of UROPs at MIT.

 \newpage
\tableofcontents
\newpage
\listoffigures
\newpage
%\listoftables
%\newpage

\pagestyle{plain}
\lfoot{\today}
\cfoot{\thepage}
\rfoot{ALFA, CSAIL, MIT}
\pagestyle{plain}
\pagenumbering{arabic}
\setcounter{page}{1}
\newpage

\section*{Preface}
This document reports on the MOOCdb project.  It will be updated as appropriate with feedback from the MOOC data science community. For latest version of this document the reader can go to the following URL: https://github.com/moocdb
 \vspace{3cm}

\noindent Version 1 - July 12, 2013. \\
Version 2 - August 15, 2013.
Version 3 - September 24, 2013 \\
Version 4 - October 15, 2013 \\

Note on contributions: The project started at CSAIL, MIT and the first version of MOOCdb data model was presented at Artificial Intelligence in Education Conference in July, 2013. Since then there have been a number of updates to the model and this technical report. The updates are a reflection of the model adaptation process. First the model was updated to incorporate additional needs of Educational researchers based on feedback from AIED. Then a set of additional attributes that edX was capturing demanded we update the model. The third round of updates reflect the adaptation of the model to the coursera platform. This adaptation required agreement on resource types, user types

 \vspace{3cm}
%\input{preface}
%\newpage

%\input{intro}
\newpage
\section{Introduction}

%Our team has been conducting research related to mining information, building models, and interpreting data from the inaugural course offered by edX, \textit{6.002x: Circuits and Electronics}, since the Fall of 2012. This involves a set of steps, undertaken in most data science studies, which entails positing a hypothesis, assembling data and features (aka properties, covariates, explanatory variables, decision variables), identifying response variables, building a statistical model then validating, inspecting and interpreting the model. In our domain, and others like it that require behavioral analyses of an online setting, a great majority of the effort (in our case approximately 70\%) is spent assembling the data and formulating the features, while, rather ironically, the model building exercise  takes relatively less time. As we advance to analyzing cross-course data, it has become apparent that our algorithms which deal with data assembly and feature engineering lack cross-course generality. This is not a fault of our software design. The lack of generality reflects the diverse, ad hoc data schemas we have adopted for each course. These schemas partially result because some of the courses are being offered for the first time and it is the first time behavioral data has been collected. As well, they arise from initial investigations taking a local perspective on each course rather than a global one extending across multiple courses.

\textcolor{black}{The ALFA Group's MOOC science} team has been conducting research using MOOC data since the Fall of 2012. Our steps, like most data science studies, \textcolor{black}{including those of our colleagues at Stanford University and Coursera}, entail positing a hypothesis, assembling data and features (aka properties, covariates, explanatory variables, decision variables), identifying response variables, building a statistical model then validating, inspecting and interpreting the model. Like others who also conduct behavioral analyses of online activities, a great majority of the effort (in our case approximately 70\%) is spent assembling the data and formulating the features, while, rather ironically, the model building exercise  takes relatively less time. 

As we advance to analyzing cross-course data and broaden our single course analysis, it has become apparent that we need a general  organization for our data which will serve multiple specific uses in these contexts.  We need to cleanup the data schemas which arose in an ad-hoc manner from each course, generalize them and take a consistent global view to data organization. If our data organization is sufficiently general to support the breadth of MOOC behavioral data analysis and extensible, it will even support cross-institute and cross-platform collaboration within MOOC data science. 

\begin{figure*}[h!]
\centering
\includegraphics [width=0.8\textwidth, height=1.4in]{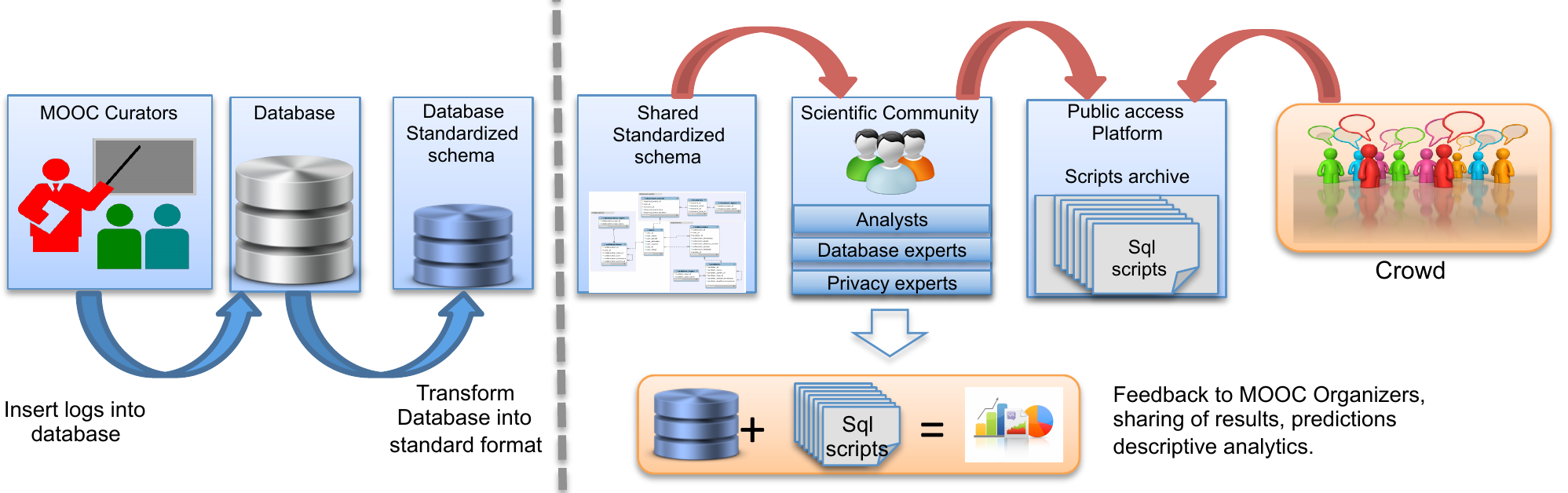}
\caption[Scenario of a standardized database schema adoption]{This flowchart represents the scenario of a standardized database schema adoption. From left to right: Curators of MOOC courses provide the raw data or format their raw behavioral interaction sources into the schema and populate either public or private databases.
% via ``\textit{piping scripts}''. The piping scripts organize and integrate the disparate streams along a standard schema adopted by the MOOC education science community. 
The analytics specialists, commonly researchers,  develop and share ``schema scripts'', based upon it. The scripts, written in standard programming languages such as Python with SQL commands embedded extract study data from schema-based database, visualizing it, conditioning it into model variables and/or otherwise examining it. The schema is unifying while the scripts support cross-institution research collaboration and direct experimental comparisons.
}\label{fig:overview}
\end{figure*}

This report documents \moocDB which is our solution to centralizing and generalizing MOOC data organization and providing general purpose analytics for MOOC education research. We offer it to the MOOC data science community at large and advocate its general adoption.  We are pleased to already have joined forces with representatives from Stanford University and Coursera. The \moocDB concept, see Figure~\ref{fig:overview},  includes a schema and shared means of scripting while avoiding any assumption of data sharing. %It supports sharing \textit{how the data is extracted, conditioned and analyzed}.  
To the extent that it can, \moocDB addresses \pii,  privacy preservation and data access control.
It also avoids some data sharing issues through the sharing of scripts which help analyze, visualize data and prepare data for models, rather than data itself while providing a means of facilitating intra and inter-platform collaboration.  It draws inspiration from similar examples of development and dissemination of well organized datasets have enabled research to thrive in medical domain\cite{saeed2002mimic} and in education domain as well \cite{koedinger2010data, stamper2010pslc}.

\newpage
% For example, it must accommodate authors publicly releasing their data assembly scripts with their publication. This will allow other researchers to replicate the reported results by executing the same data assembly script on their own data. Comparisons will be accurate and direct because any sensitivity and variance due to data conditioning will be eliminated. 
 
Sharing scripts which analyze, visualize data and prepare data for models, rather than data itself, will not only help mitigate privacy concerns but it will also provide a means of facilitating intra and inter-platform collaboration.  One goal of \moocDB is to set a standard facilitating insights from data being shared without  data being exchanged. For example, two researchers, one with data from a MOOC course on one platform and another with data from another platform, should be able to decide upon a set of variables, share ``\textit{schema}'' scripts that can extract them, each independently derive results on their own data, and then compare and iterate to reach conclusions that are cross-platform as well as cross-course. 

\moocDB will also enable research authors to release a method for recreating the variables they report using in their published experiments. It will also enable community driven contribution of open source analytics software. For example, a computer scientist interested in this data can optimize a schema script to extract data at really fast speeds and contribute it to the publicly available, open source, library. At the same time, an educational science researcher can download schema scripts and execute them on their data reaping the benefits of the cross-disciplinary collaborative platform.  
 %is subsequently analyzed through \textit{``assembly scripts''}.
%
%Assembly scripts will enable research authors to release a method for recreating the variables they report using in their published experiments. We envision that authors will publicly release their data assembly scripts with their publication, likely on a project website. This will allow other researchers to replicate the reported results by executing the data assembly script on their own data. Comparisons will be accurate and direct because any sensitivity and variance due to data conditioning will be eliminated. Sharing scripts which prepare data for models, rather than data itself, will not only help mitigate privacy concerns. It will also provide a means of facilitating intra and inter-platform collaboration. Two researchers, one with data from a MOOC course on one platform and another with data from another platform, could decide upon a set of variables, each independently derive results on their own data, and then compare and iterate to reach conclusions that are cross-platform as well as cross-course. In a practical sense, insights from data are being shared without  data being exchanged.

Our contention is that the MOOC data mining community - representing all branches of educational research as it relates to online learning, need to act immediately to engage in consensus driven discussions toward a means of standardizing data schema and building complimentary technology that enables collaborating on data science via sharing scripts, resulting in a practical, scalable solutions to data science in this domain. It is important to take initial steps now. We have the timely opportunity to avoid the data integration chaos that has arisen in fields like health care where large legacy data, complex government regulations and personal privacy concerns are starting to thwart scientific progress and stymy access to data.  We hope this report and \moocDB contribute to this process. 
% \footnote{We would like to use the MOOCshop  as a venue for introducing it and offering it up for discussion and feedback. We also hope to enlist like minded researchers willing to work on moving the concept forward in an organized fashion, with plenty of community engagement.}

%Our contention is that the MOOC data mining community - from all branches of educational research, should act immediately to engage in consensus driven discussions toward a means of sharing results in a practical, directly comparable and reproducible way. It is important to take initial steps now. We have the timely opportunity to avoid the data integration chaos that has arisen in fields like health care where large legacy data, complex government regulations and personal privacy concerns are starting to thwart scientific progress and stymy access to data.  In this contribution, we propose a standardized, cross-course, cross-platform, database schema which we name the \textit{Education Science Unifying Database  Schema} or \textit{``EDSCUDS''} \footnote{we do not like this name yet-UMO and KV}.   We would like to use the MOOCshop  as a venue for introducing it and offering it up for discussion and feedback. We also hope to enlist like minded researchers willing to work on moving the concept forward in an organized fashion, with plenty of community engagement. 

We proceed by describing the concept of \moocDB and what it offers in more detail in Section~\ref{sect:concept}.  Section~\ref{sect:challenges} outlines \moocDB challenges.  Section~\ref{sect:schemaDescription} details our proposed the data schema  systematically. Section~\ref{sect:useCase} shows, with a use case, how the current data storage mechanisms and ad-hoc analytic frameworks are not scalable. Section~\ref{sect:tools} presents multiple tools we built that are enabled by the schema and how the schema is expressive, supportive and reusable. Section~\ref{sect:conclusions} concludes and introduces our current work.

%We outline three categories of data features; course, student, and course-student interaction features and provide a schema and example features for the course-student category. Future work will devise schemas for all categories and provide a library of features and their generating scripts for use in the schema. 

%\input{Related}
\section{MoocDB: Its Concept and Benefits}\label{sect:concept}

\moocDB:
\begin{itemize}

\item identifies two kinds of primary actors in the MOOC eco-system: \textit{curators} and \textit{analysts}. Curators archive raw behavioral data expressing MOOC students' interaction with online course material, often storing it in multiple formats and streams. This data can be passed to researchers as a collection of file dumps of databases, logs, json or other formats. Curators include course content providers and course platform providers.  \textit{Analysts} reference the data to examine it for descriptive, inferential or predictive insights. The role of the analysts is to visualize, descriptively analyze, and use machine learning or otherwise interpret some set of data from the source set.  Analysts  extract, condition (e.g. impute missing values, de-noise), and create higher level variables for modeling and other purposes from the data. To perform this analysis, they first transform the data by organizing into the standard schema and compose \textit{scripts} or use publicly available scripts when that suffices. They also contribute their scripts to the archive for use by others.

\item identifies two types of secondary actors: the \textit{crowd}, and the information technology specialists with expertise in databases, data access and security. When needs arise, the community can seek the help of the \textit{crowd} in innovative ways. Experts contribute to the community by providing state-of-the art technological tools and methods.

%\item Curators independently format the raw data, \textit{using a schema}, to populate a database repository. Alternatively, the data may be directly collected into a database according to a schema.
\item Includes a common standardized and shared schema into which the raw data is subsequently organized and structured. \textcolor{black}{The schema organizes the information in terms of 4 different behavioral interaction modes: \begin{enumerate}
\item submitting
\item observing
\item collaborating 
\item feedback
\end{enumerate}
It organizes user identification information in a manner that allows different privacy preserving options.} Our goal is that the schema becomes agreed upon by the community, generalizes across platforms and preserves all the information needed for data science and analytics. 
%\item The database repository can be public or private depending on the access policy of its curator.
\item includes a  shared community-oriented repository of data extraction, feature engineering, and analytics scripts all based on the common data schema. \textcolor{black}{In this report, (see Section~\ref{sect:MoocEnImages}) we describe MoocEnImages, an open source, open ended framework we intend to release in the coming year.}

\item over time, supports the open-ended growth of the repository  and evolution of the schema. 
%\item 
%\item The community of analysts are able to reuse assembly scripts for further analysis, comparative analysis or collaborative analysis. 
\end{itemize}

\noindent \moocDB offers:
\begin{description}

\item \textbf{The benefits of standardization}: The data schema standardization implies that the raw data from every course offering will eventually be formatted the same way.   It establishes simple conventions like ``storing event timestamps in the same format'' to common tables, fields in the tables, and relational links between different tables. It makes it possible to populate a database, which because of its organization can appropriately support scientific discovery and inquiry.   Standardization supports cross-platform collaborations, sharing query scripts, and the  definition of variables which can be derived in exactly the same way  irrespective of which MOOC database they come from.

\item \textbf{Concise data storage}: \moocDB{'s} proposed schema  is ``loss-less'' with respect to research relevant information, i.e. no information is lost in translating raw data to it. However, the use of multiple related tables provides more efficient storage.  Some information, such as fields specific to a provider that is not generalizable and irrelevant to researchers is culled.%Encoding schemes that reduce the size of the raw data by employing dictionaries, so that we can load it in raw memory. 
% I am not sure we want to claim that our data is loss-less in this respect. It my remove data that we feel is irrelevant to researchers or very course/provider specific fields.

\item \textbf{Access Control Levels for Anonymized Data}:  The data schema offers an organized means of structuring anonymized user identities  safeguard them further. It provides 3 different levels of  assurance via   cross-referencing information across 3 tables. By controlling which tables are released, a different level of assurance is available. See Section~\ref{sect:usertable} for details.

\item \textbf{Savings in effort}:  
A schema speeds up database compilation by eliminating repeated schema design.  Investigating a dataset using one or more existing scripts helps speed up research. 

\item \textbf{Sharing of data extraction scripts} Scripts for data extraction and descriptive statistics extraction can be open source and shared by everyone because they reference data organized according to the schema. Some of these scripts could be very general and widely applicable, for example: ``For every video component, provide the distribution of time spent by each student watching it" and some would be specific  for a research question, for example generation of data for Bayesian knowledge tracing on the problem responses. These scripts would evolve to become optimized by the community and perpetually updated. 

\item \textbf{Crowd source potential}: Machine learning frequently involves humans identifying explanatory variables that could drive a response. Enabling the crowd to help propose variables could greatly scale the community's progress in mining MOOC data. We intentionally consider the data schema to be independent of the data itself so that people at large, when shown the schema, optional prototypical synthetic data and a problem, can posit an explanatory variable, write a script, test it with the prototypical data and submit it to an analyst. The analyst can assess the information content in the variable with regards to the problem at hand and rank and feed it back to the crowd, eventually incorporating highly rated variables into learning.  

\item \textbf{A unified description for external experts} For experts from external fields like``Very Large Databases/Big Data" or "Data Privacy", \moocDB's standardization presents data science in education as unified.  This allows them to technically assist us with techniques such as new database efficiencies or privacy protection methods. 

\item \textbf{Sharing and reproducing the results}: When they publish research, analysts can share the scripts by depositing them into a public archive where they are retrievable and cross-referenced to their donor and publication.  This allows results to be tested on additional data with precisely the same methods.

\end{description}

\section{Challenges}\label{sect:challenges}
\moocDB faces the following challenges: 

\begin{description}
\item \textbf{Schema adequacy:} The \moocDB standardized schema must express all the information contained in the raw data, now and in the future as MOOC platforms evolve. To date, we have verified it with multiple courses and across the edX and Coursera platforms. We expect the schema to significantly change as more courses and offerings are explored. It will be challenging to keep the schema open ended but not verbose. While a committee could periodically revisit the schema and make collective decisions on how to update it, a more robust approach would be to let it evolve in the course of being extended and used.  Good extensions will be adopted frequently and can become permanent. Lesser ones will fade from usage and can be retired. This strategy would recognize the diversity and current experimental nature of MOOC science and avoid  standard-based limitations.

One example of a context similar to the growth of MOOCs is the growth of the internet. HTML and Web3.0 did not attempt rein in the startling growth or diversity of world wide web components. Instead, HTML (and its successors and variants) played a key role in delivering content in a standardized way for any browser.  As the browser and backend computing evolved,  web standards extended and grew with retirement of redundant parts. At this time,  the semantic web provides a flexible, community driven, means of standards adoption rather than completely dictating static, monolithic standards. 

\item \textbf{Could \moocDB play a role in the breach of private data?} 
It is imperative to make all reasonable efforts to afford appropriate privacy to all students in MOOCs. However there are always determined people who marshall complex technical means to try to de-identify the data over the will of people who make every effort to anonymize it via mechanisms such as aggregation or noising.
  
By its current level-based organization of anonymized user identification information, \moocDB can be used to reduce risk of re-identification, given anonymization. However it poses several privacy breach risks: \begin{itemize}
\item Forum data could be used to identify a user within the MoocDB data
\item Outside data could be used to identify a user within the MoocDB data
\item MoocDB data could be used, even though anonymous, to  identify a user in data outside the MoocDB data model
\end{itemize} 

One related challenge is the \moocDB must function in the absence of clear data control policies.  Local policies at the institution or platform level are still in flux or in need of formalization. These would allow MoocDB to better address the issue of privacy preservation.

\item \textbf{Platform Support}: The community needs a website hosting the \moocDB schema definition. It requires a means of assisting researchers in sharing scripts. It requires tests for validating scripts, metrics to evaluate new scripts and an repository of scripts with efficient means of indexing and retrieval. 

\item \textbf{Motivating the crowd}: We intend to use \moocDB's standard schema to help us use the crowd as a source of defining covariates for statistical analysis. The challenge of using a crowd to suggest covariates is how to describe the data and provide useful feedback without granting or controlling access to the data itself.  Here, KAGGLE provides a framework from which we can draw inspiration. KAGGLE provides a problem definition plus a dataset that goes along with it. Using \moocDB we can also propose a problem. But, in contrast to KAGGLE we will share relevant elements of the \moocDB schema rather than pass a dataset. We will be able to provide an example of a set of covariates and the scripts that enabled their extraction. This will allow us to encourage users to posit covariates and submit scripts that extract them from the data.  %Such an endeavor requires us to: provide synthetic data (under the data schema) to allow the crowd to test and debug their scripts, define metrics for evaluation of covariates/features given the problem,  and possibly visualizations of the features or aggregates over their covariates (when possible), and most importantly a dedicated computing resource that will perform machine learning and evaluate the metrics over which we evaluate the information content in the user provided covariates. 

\end{description}

\vspace{-3mm}

\section{Schema description} \label{sect:schemaDescription}

We surveyed a typical set of courses from Coursera and edX and observed three different modes in which students engage with the material. First, students observe the material by accessing all types of resources. In the second mode, they submit material for evaluation and feedback. This includes problem check-ins for lecture exercises, homework and exams. The third mode is in which they collaborate with each other. This includes posting on forums and editing the wiki. In the future it might include more collaborative frameworks like group projects. We name these three modes as \textit{observing}, \textit{submitting} and \textit{collaborating}.  Based on these we divide the database schema into three sets of different tables. We now present the data schema for each mode. As noted before, the schema in its entirety does not eliminate any relevant information in the raw data. Note that some tables include some columns which may be optional.

%%%%%%% Observing

\subsection{The observing mode}

In this mode, students simply browse and observe a variety of resources available on the website. These resources include the \textit{wiki}, \textit{forums}, \textit{lecture videos}, \textit{homeworks}, \textit{book}, \textit{tutorials}. Each unique resource is usually identifiable by a unique resource identifier \textit{uri}.  A single webpage, identified by an \textit{url}, can contain multiple \textit{uri}s that might not be visible to the user, but his/her access is recorded in the data as an access to the specific \textit{uri} within the \textit{url}. For the case where there is only resource on a webpage, the \textit{url} and \textit{uri} are the same. Additionally, $uri$s can occur on multiple $urls$.  Through our schema we capture: the exact resource which the user accessed, and when available the context in which the resource was accessed by the user. 

\begin{figure}[htb!!]
\centering
\includegraphics [width=0.8\textwidth]{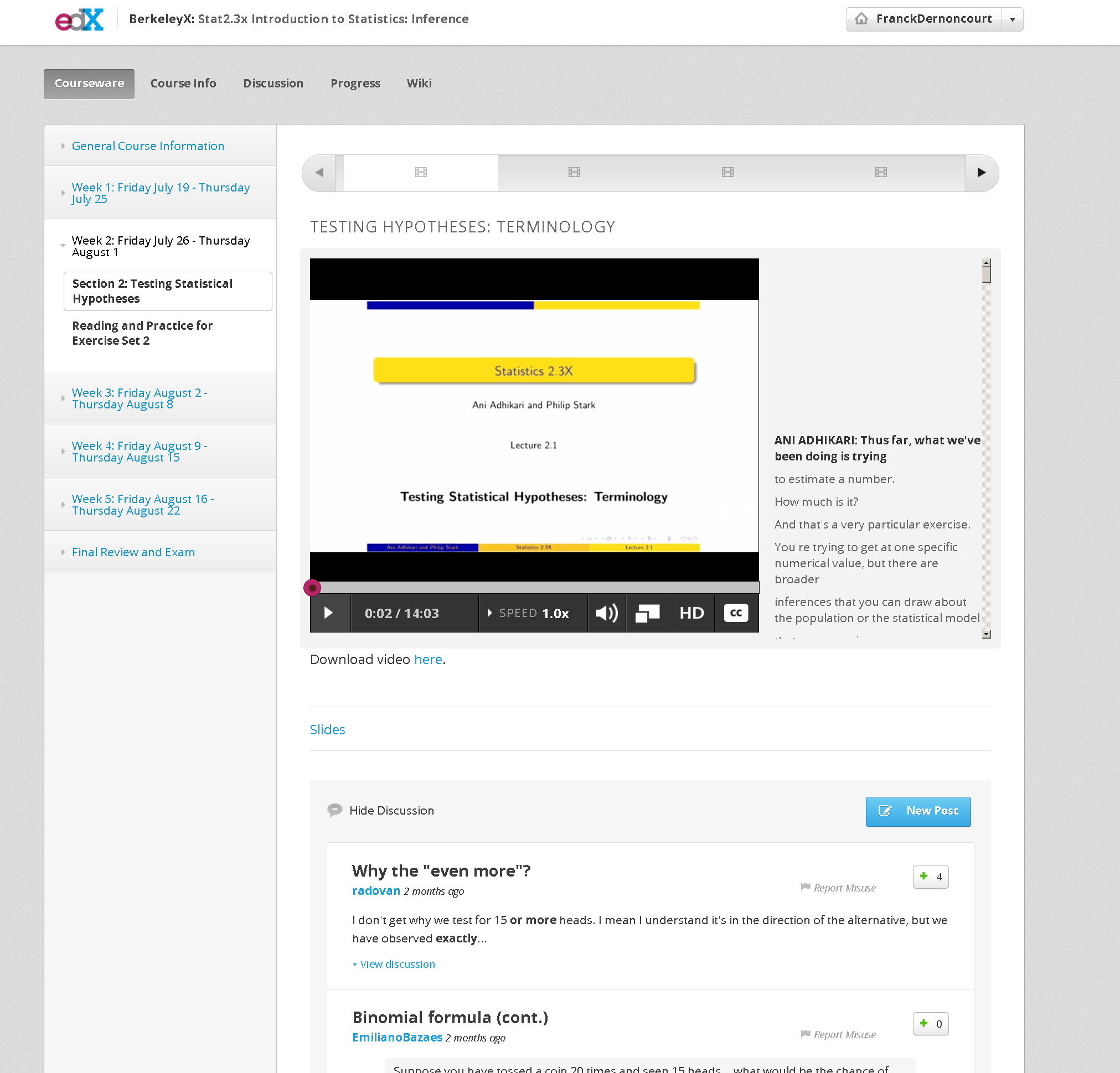}
\caption[Example of multiple resources on the same url]{An example from an edX course where multiple resources are visible on the same \textit{url}. In this page both lecture video and forum posts corresponding to the lecture video are presented to and accessed by the students.}
\label{fig:Observing}
\end{figure}

We propose that data pertaining to the observing mode be formatted in 5 tables: a transaction based observed events table, resources table, a dictionary table that identifies the resource type, a \textit{url}s table and a $resource\_urls$ tables. Figure~\ref{fig:Observing} shows the schema and the links. 

\begin{description}
\item \textbf{Observed\_events table}: This holds all observed events for a student: $observed\_event\_id$, $resource\_id$,  $user\_id\_observed$, $observed\_event\_duration$, $observed\_event\_timestamp$, $observed\_event\_ip$, $observed\_event\_os$, $ob\-served\_event\_agent$. Each row corresponds to a student's visit to a resource URI.  The $url\_id$ stores the url where the particular resource was accessed by the user. This transaction-based table is sufficiently generalizable to record any type of a user's resource usage patterns. 

\begin{figure}[htb!!]
\centering
\includegraphics [width=0.6\textwidth]{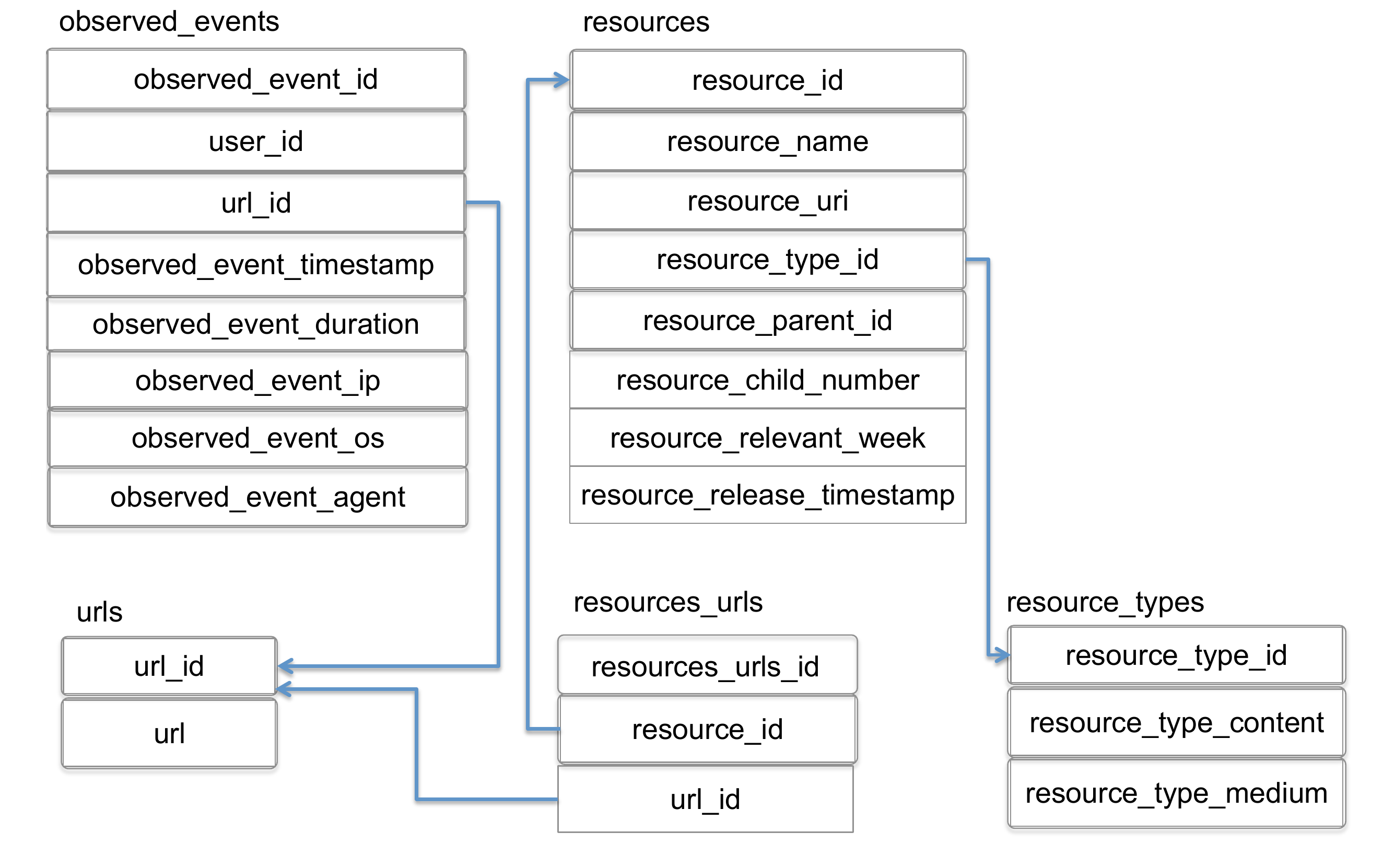}
\caption[Data schema for the observing mode]{Data schema for the observing mode. Note that for readability, the field names may be simplified and the diagram may not include every field. See Figure~\ref{fig:FullSchema} for the full schema.}
\label{fig:Observing}
\end{figure}

\item \textbf{Resources table}: This dictionary table maps a \textit{uri} to a $resource\_id$. Its fields are $resource\_id$, $resource\_name$, $resource\_uri$, $resource\_parent$ and $resource\_type\_id$ and a $resource\_child\_number$ which stores the order of the resource if it shares a parent with another resource in the resource hierarchy. 

\item \textbf{urls table}: This gives a unique number, $url\_id$ for each url on the website. 

\item \textbf{Resource\_types table}:  This dictionary table maps $resource\_type\_id$ to a resource type. The set of resource types are, $\{$\textit{book}, \textit{wiki}, \textit{forums}, \textit{exercises}, \textit{video}, \textit{problems}, \textit{tutorials}, \textit{lecture}$\}$

%\item \textbf{Users table}:  This table contains basic user data, such as his or her id and name. It is common across all three modes, but is introduced here. Its column fields are: $user\_id$, $user\_name$, $user\_gender$, $user\_birthdate$, $user\_country$, $user\_ip$, $user\_timezone\_offset$, $user\_final\_grade$, $user\_join\_timestamp$, $user\_os$, $user\_agent$, $user\_language$, $user\_screen\_resolution$. The timezone offset is the offset with GMT zone, which is used to put timestamps in user's timezone if needed. The final grade is on a 0-1 scale, and is normalized if necessary. 

\end{description}

%%%%%%%%%%% Submitting

\subsection{The submitting mode}

Similar to the tables for the observing mode of the student, we now present  a structured representation for the data that records student interactions with the assessment modules of the course. In this mode the student submits a response/answer to a problem/question in the course and receives feedback and evaluation. We call this \textit{submitting} mode. A typical MOOC student is in the submitting mode when trying homework problems, exams, quizzes, exercises in between lectures and labs (for engineering and computer science). 

In contrast to submissions in campus settings, MOOC submissions are more complex. While question types rarely differ much, that is they can be multiple choice or require an analytical answer, a program or an essay, submission software enables students to submit answers and check them multiple times, save their intermediate work and submit later. Some courses limit the maximum number of submissions per question for either all or some subset of questions \cite{genesereth13, koller13}. The homeworks can have soft and/or hard deadlines \cite{genesereth13} resulting in data indicating what deadline a student met. 

Due to the online nature of submissions, assessments are handled in different ways.  Assessments could be done by the computer via simple check mechanisms or automated algorithms \cite{6.002xfirst}, peer review \cite{klemmer13}., evaluation by instructors and/or graders. For some courses multiple assessors are used \cite{piech13}. We design the schema to be able to capture this information. 

In our experience, submissions and their evaluation mechanisms in MOOCs are far more complex and require more complex organization. To express all the nuanced information, we designed three tables, the \textit{problems} which captures the problem structure, the \textit{submissions} which is transactional and stores every submission by every student, and the \textit{assessments} tables which stores the evaluations and feedbacks.  Figure~\ref{fig:Submitting} shows the schema and the links for this mode. 
 The tables in this mode and their cross reference links are:  
\begin{description}
\item \textbf{Problems table}: This table stores information solely about the problems.  An important aspect of this table is that it captures the problem hierarchy (i.e. a problem, a subproblem, its subproblem etc.). The schema identifies each problem module, a quiz or a homework, with a tree hierarchy with the root node identifying the module it self. The leaves of this tree represent the sub problems to which a student submits an answer to. Each node in the tree is given a unique number. For example, in Figure~\ref{fig:phierarchy} we show homework 1. The root is homework 1 and homework 1 has two problems. The first problem has 4 sub problems and the second has 3 sub problems. The nodes are numbered as shown in the figure. Unique numbers are generated for every module represented as a tree. 
\begin{figure}[ht]
\centering
\includegraphics [width=0.45\textwidth]{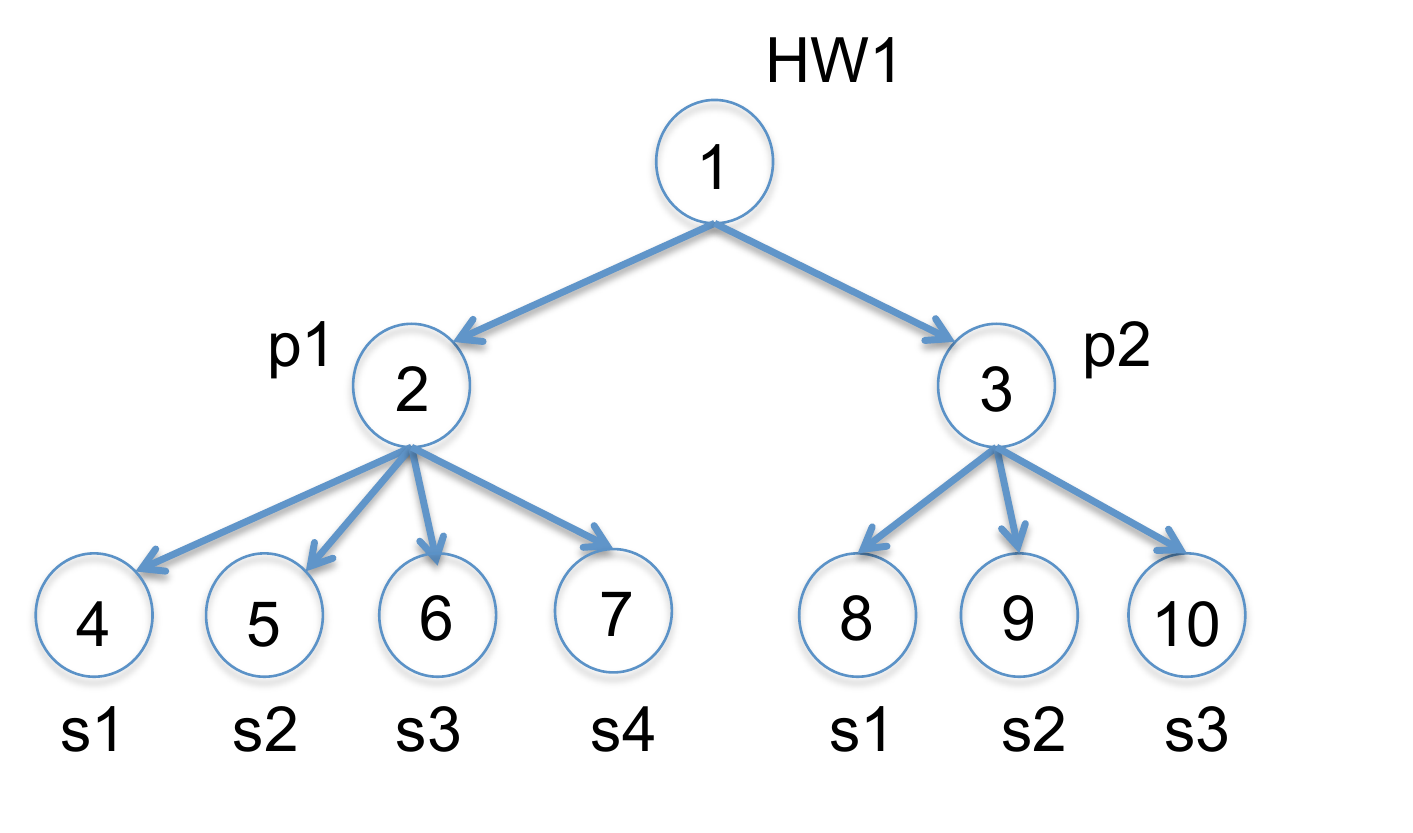}
\includegraphics[width=0.45\textwidth]{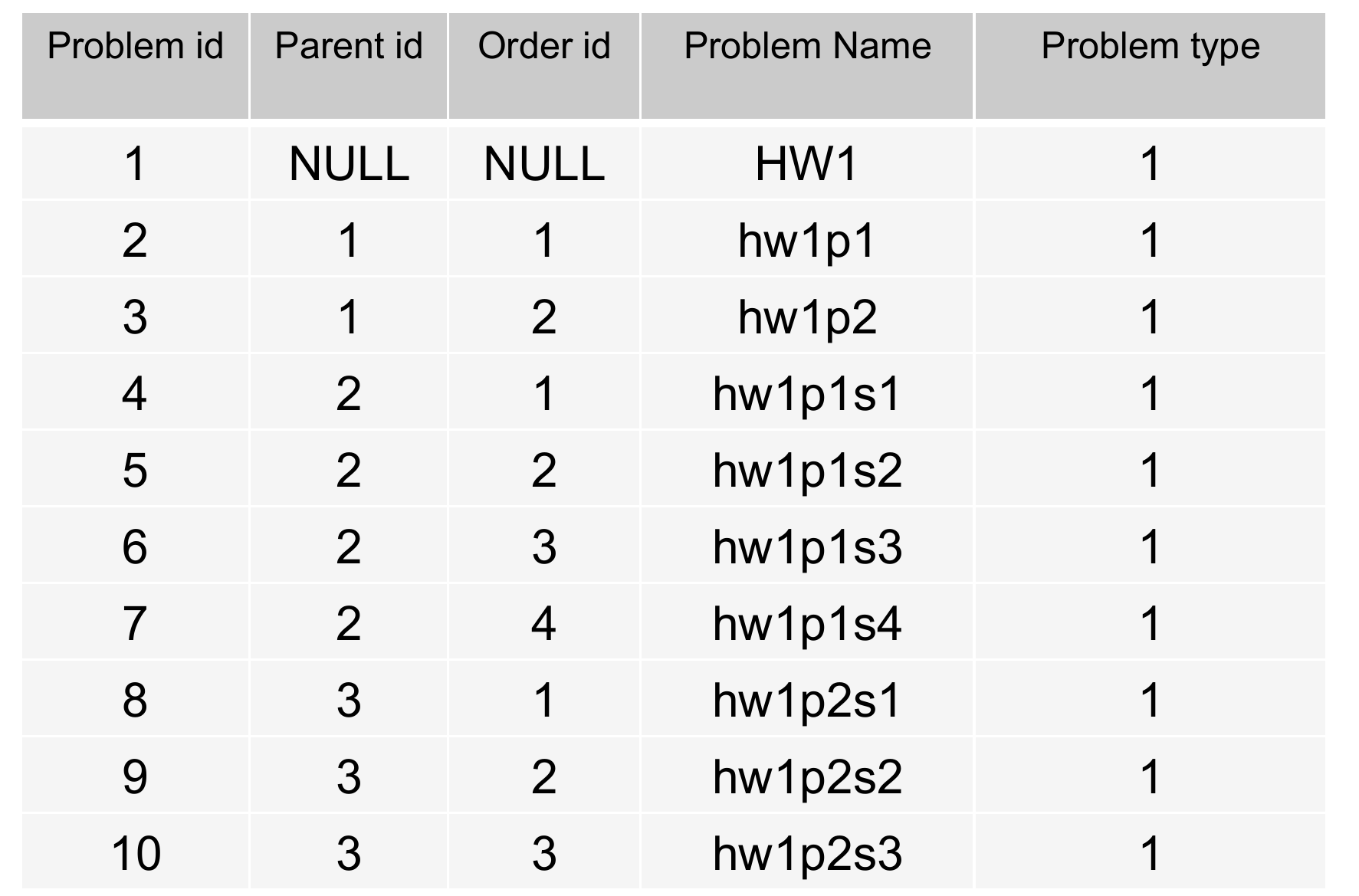}
\caption[Data schema for the submitting mode]{Representing the hierarchy of problems within a module as a tree. In this example, homework 1 has two problems which have 4 sub problems and 3 sub problems respectively. Each node in the tree is identified by a unique number which becomes is recorded as $problem\_id$. On the right, the corresponding entries in the table are shown. Note that only a subset of fields in the table relevant to the example are shown.}
\label{fig:phierarchy}
\end{figure}

Each node in the tree has a corresponding entry in the table. The number of the node is its $problem\_id$. The schema stores the hierarchy information by recording a problem's `parent' as a $problem\_parent\_id$. A problem with no `parent' will have this field as a null.  Multiple subproblems are also given $order\_id$ preserving their order with in the problem. This information allows the researcher to fully reconstruct the topology of the questions if needed. 

Additional information regarding the problem is stored in this table are: $problem\_type\_\-id$ which captures whether it is a homework, quiz, midterm or lecture exercise, $problem\_name$, $problem\_release\_timestamp$, $problem\_\-hard\_dead\-line\_timestamp$, $problem\_\-soft\_deadline\_timestamp$, and $problem\_max\_submission$. 

%The table consists of: $problem\_id$, $problem\_name$, $problem\_parent\_id$, $order\_id$, $problem\_type\_\-id$, 
%The schema allows for multiple deadlines, which are used by some MOOC providers. 

\item \textbf{Problem\_types table}: This table is meant as a dictionary to differentiate the different types of problems, such as a homework, exercise, midterm or final.  It includes the $problem\_type\_id$ and $problem\_type\_name$ fields.

\item \textbf{Submissions table}: In this table each submission made by a student is recorded. Each entry in the table corresponds to a submission made by the student recorded as: $submission\_id$, $user\_id$, $problem\_id$, $submission\_timestamp$, $submission\_answer$, $submission\_attempt\_\-number$, $submission\_ip$, $submission\_os$, $submission\_ agent$ and $is\_submitted$. The $is\_submitted$ field allows for courses to include submissions that are not graded yet (such as a draft). 

\begin{figure}[htb!!]
\centering
\includegraphics [width=0.6\textwidth]{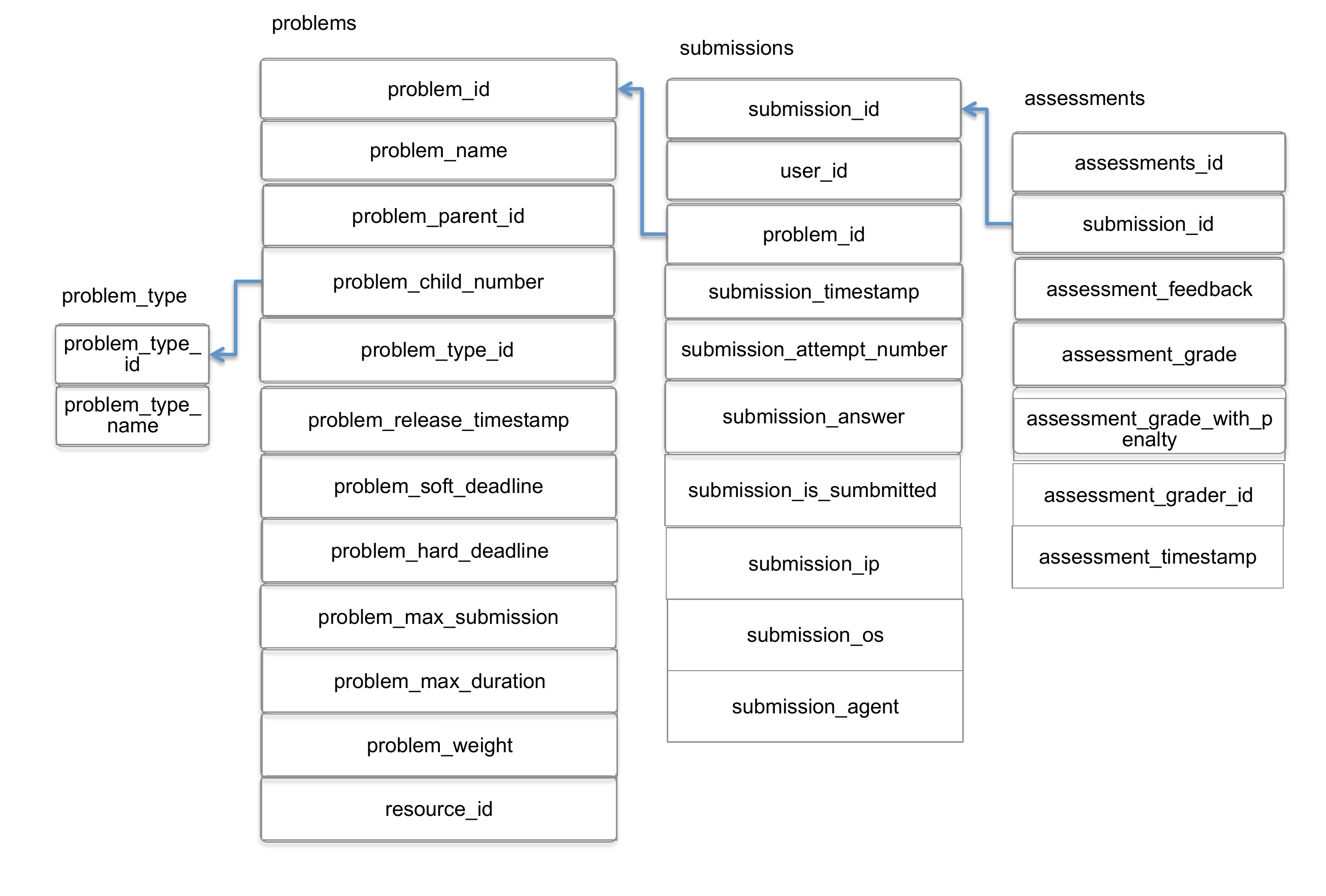}

\caption[Data schema for the submitting mode]{Data schema for the submitting mode. Note that for readability, the field names may be simplified and the diagram does not include the user table. See Figure~\ref{fig:FullSchema} for the full schema.}
\label{fig:Submitting}
\end{figure}

\item \textbf{Assessments table}: This table decouples the assessment from its submission, which allows for multiple assessments per submission. In this table each assessment for a submission is stored as a separate row. The table consists of:  $assessment\_id$ and $submission\_id$, $assessment\_feedback$, $assessment\_\-grader\_id$ (the \textit{user\_id} of the grader), $assessment\_grade$, and $assessment\_\-timestamp$. This schema allows for multiple graders for a submission. Also, the grade is meant to be normalized (the maximum grade is always a 1).

\end{description}
\vspace{-3mm}

%%%%%%%%%% Collaborating

\subsection{The collaborating mode}

Students collaborate among themselves throughout the course through mediums such as a forum and a wiki. In a forum, a student either initiates a new thread or responds to an existing thread. Additionally students can `up vote', and `down vote' the answers from other students. In a wiki, students edit, add, delete and initiate a new topic. Although the method of collaboration differs, different  student-to-student interactions can be captured with a single record system. This generalizability also allows for future types of student collaborations. Another large advantage of this method of recording wiki and forum data is that the log is per-student and transaction based. This allows researchers to more easily get data about a student, rather than a thread or wiki page. Figure~\ref{fig:Collaborating} shows the schema and the links. To capture this data we propose the following tables and fields: 

\begin{figure}[htb!!]
\centering
\includegraphics [width=0.6\textwidth]{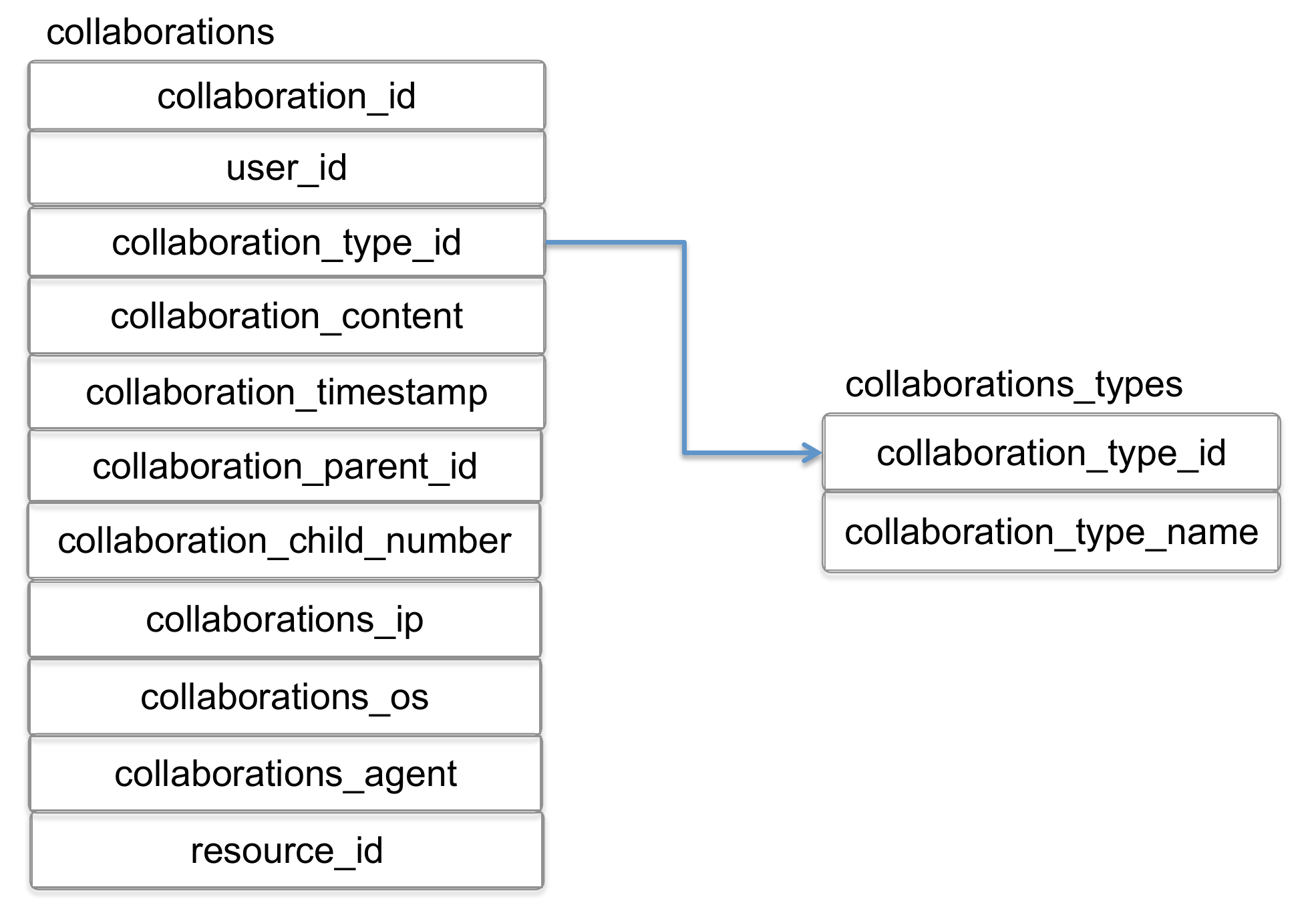}
\caption[Data schema for collaborating mode]{Data schema for collaborating mode. Note that for readability, the field names may be simplified and the diagram may not include every field. See Figure~\ref{fig:FullSchema} for the full schema.}
\label{fig:Collaborating}
\end{figure}
\begin{description}
\item \textbf{Collaborations table}: In this table each attempt made by a student to collaborate is recorded. The fields in this table are $collaboration\_id$, $user\_id$,  $collaboration\_type\_id$, $collaboration\_timestamp$, $collaboration\_\-content$, $collaboration\_ip$, $collaboration\_os$, $collaboration\_agent$, and $collaboration\_parent\_id$. A collaboration's parent id is the $collaboration\_id$ of the `parent' collaboration, whose meaning is dependent on context. However, like the problems table, this allows to fully conserve the topology. To record a response on the forum, for example, a the post would be the collaboration. Its parent might be the question (also a collaboration) that the response replies to. Alternatively, a parent could be the collaboration that a student is deleting on a wiki, or the comment on a forum the student is voting on. Of course, many collaborations might not have a parent (and result in a null field). The $collaboration\_content$ is text whose value is dependent on the type of collaboration. It could be simply the post, or JSON formatted text with a question and a title.

\begin{figure}[htb!!]
\centering
\includegraphics [width=0.6\textwidth]{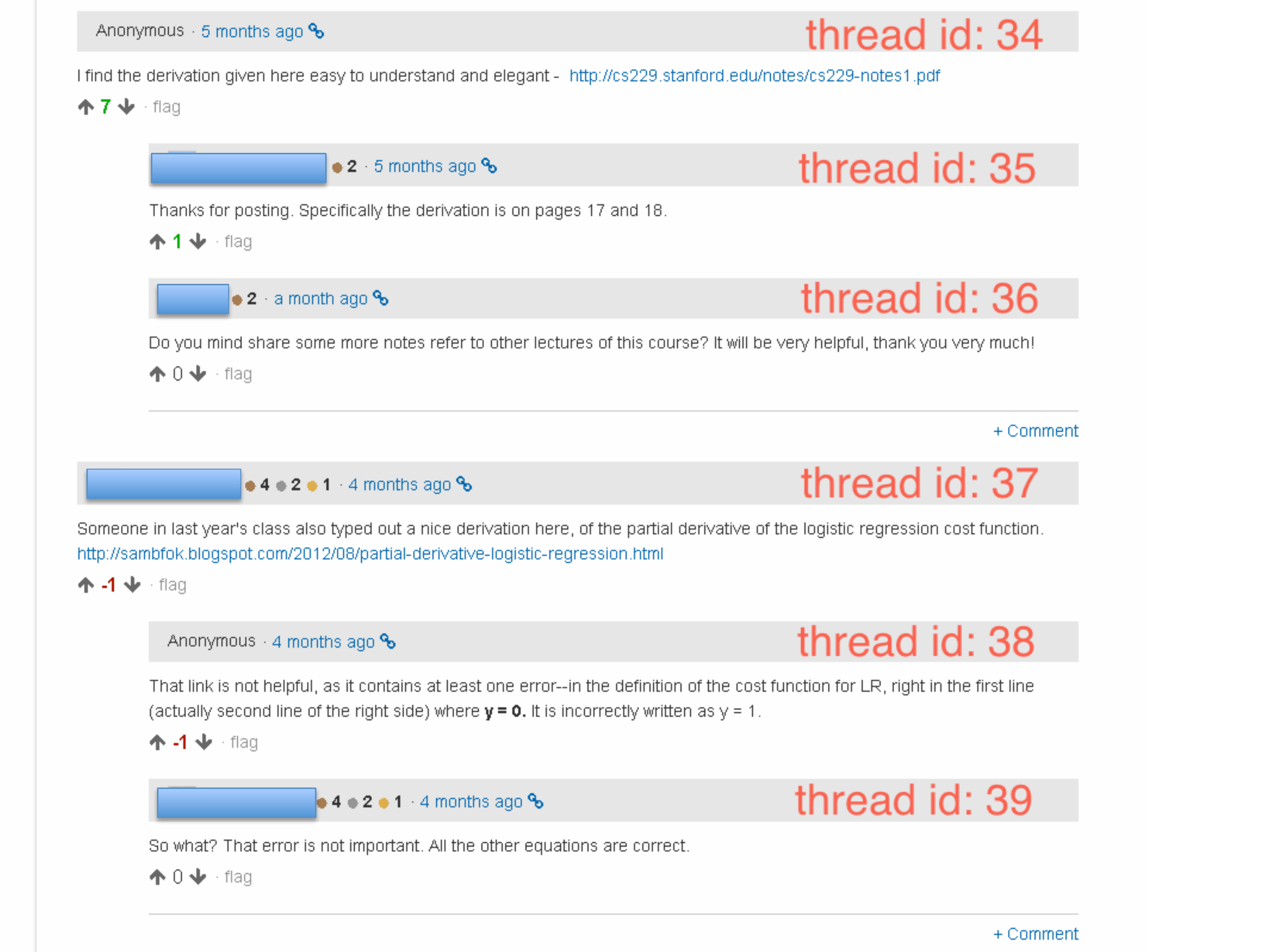}
\includegraphics [width=0.6\textwidth]{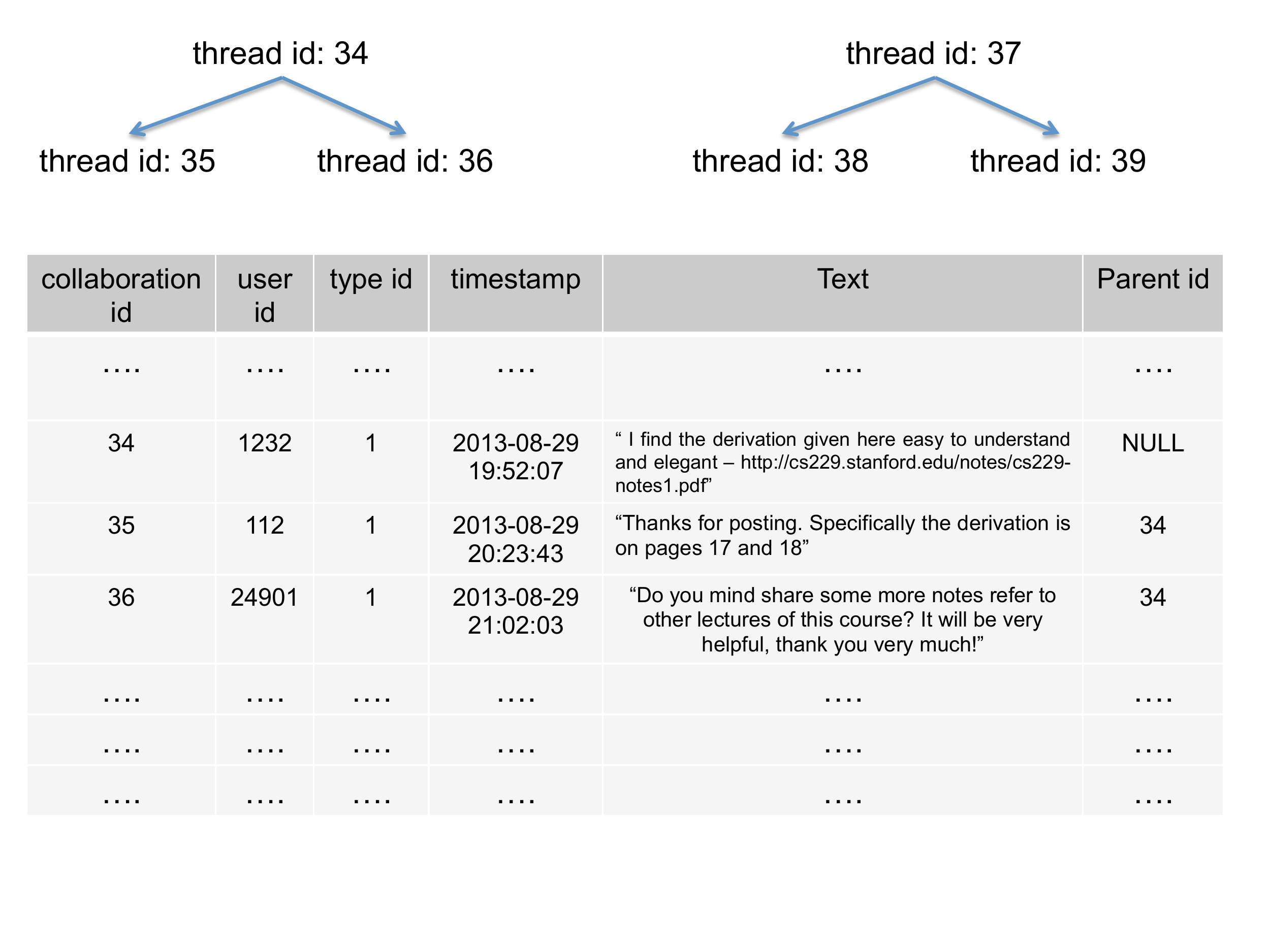}
\caption[Example of data in the collaborating mode]{Top: Shows an example set of forum posts from a coursera course. The thread is initiated by a user to whom two other users respond with comments. Bottom: This shows the thread hierarchy and populating the table in the database for the first set of posts.}
\label{fig:CollaboratingExample}
\end{figure}

\item \textbf{Collaboration\_types table}: This table is a dictionary describing different types of collaborations. Possible types could be wiki posts, wiki deletions, forum questions, forum comments, forum votes, forum replies, etc. This table includes $collaboration\_type\_id$ and $collaboration\_type\_name$.

\end{description}

%%%%%%%%%% Feedback

\subsection{The feedback mode}

This final mode captures student feedback to MOOC providers or lecturers. Often, this may be in the form of a survey response or a course rating, but as in other modes, the schema is meant to be flexible enough to allow for other forms of student to provider interactions. Figure~\ref{fig:Feedback} shows the schema and the links. To capture this data we propose the following tables and fields: 

\begin{figure}[htb!!]
\centering
\includegraphics [width=0.6\textwidth]{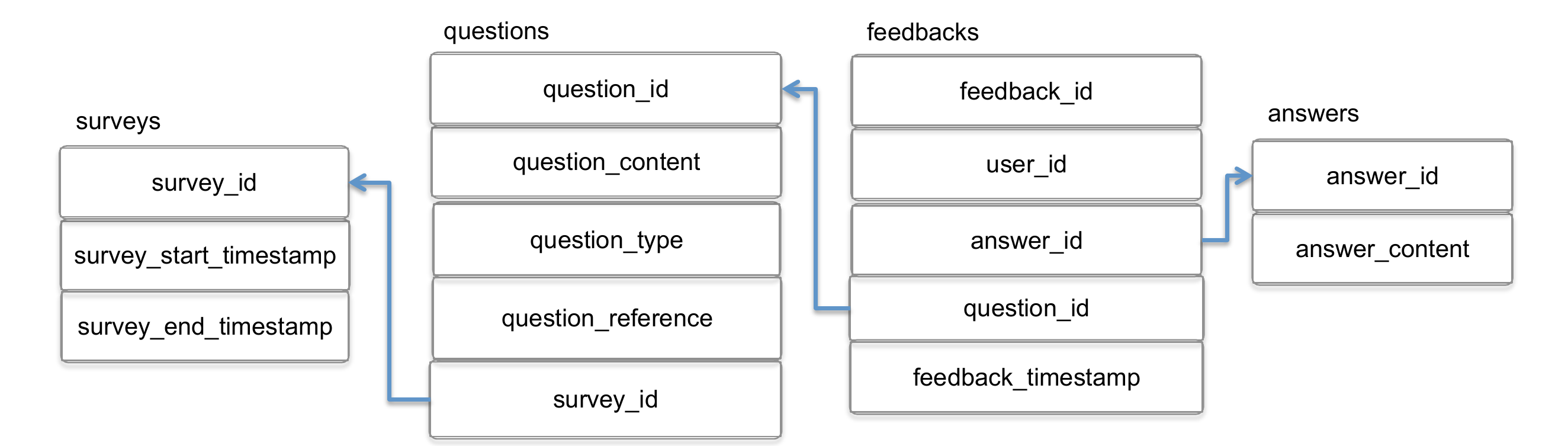}
\caption[Data schema for feedback mode]{Data schema for feedback mode. Note that for readability, the field names may be simplified and the diagram may not include every field. See Figure~\ref{fig:FullSchema} for the full schema.}
\label{fig:Feedback}
\end{figure}

\begin{description}
\item \textbf{Feedbacks table}: In this table each feedback by a student is recorded. The fields in this table are $feedback\_id$, $user\_id$,  $answer\_id$, $question\_id$ and $feedback\_timestamp$. Thus, each feedback from a student essentially consists of a (optional) question and an answer, who's id's are recorded here. The answer is the feedback, and the question is what the feedback is in response to.

\item \textbf{Questions table}: This table contains the questions that a student's answer is in response too. Examples include a survey question or the topic of what is being rated. The fields are $question\_id$, $question\_content$, $question\_type$, $question\_reference$ and $survey\_id$. $question\_content$ could be used to record the text. The $question\_reference$ contains the $resource\_id$ of the resource that the question applies to. The $survey\_id$ can be used to group questions together into a survey.

\item \textbf{Answers table}: This table is a dictionary containing student's actual feedback and the id associated with the answer. It's fields are $answer\_id$ and $answer\_content$.

\item \textbf{Surveys table}: This table is used to define a survey, which is a group of questions. It's field's include $survey\_id$, $survey\_start\_timestamp$ and $survey\_end\_timestamp$.

\end{description}

\section{User Information}\label{sect:usertable}
\newcommand{\id}{identifier\xspace}
\vspace{-3mm}

A final set of tables in MOOCdb structure anonymous, access controlled personal information concerning MOOC users. They decrease the likelihood that the anonymized information will be used to re-identify a student or group.  The \moocDB \textit{user PII table} has entries which each describe somewhat standard demographic information about the user, that is his/her age, country, most frequently used IP address. An additional field is an \id called ``\textit{global user  id}". This table should require control authorization to be shared. It should reside, encrypted, at a secure, isolated location. The ``\textit{global user id}" is an index that aids anonymization. It is also available outside the \textit{user PII table} in a second table named the \textit{global user table}. Here it cross-references to multiple entries where each is a unique  \textit{course\_user\_id}, one per course a user has taken. With it, the identifiable information can be retrieved. The decomposition of the \textit{global user  id} into one \textit{course\_user\_id} per course allows control over whether a user is able to be linked across courses. This control may be granted or denied by providing or not providing access to the  \textit{global user table}.

Finally, we also decompose the \textit{course\_user\_id} into a unique ``\textit{mode user\_id}'', one per mode (observed, submissions, collaborations, feedback). This information is in the ``\textit{course user table}'' which also contains the user's final grade in the course, its type and optional information non-\pii.   In this manner, access allowing one user to be tracked across the different modes of interaction, within a single course, is controlled by granting access to the \textit{course user table}.

%One can protect privacy via a few different approaches. The first approach allows researchers to only gain access to the aggregates. In the second approach a synthetic data generated from statistical models that emulate the original data is released for research purposes. In the third approach the personally identifiable information is masked by adopting a few strategies that could reduce the amount and type of data shared. We designed the schema to allow flexible access to the information with different levels of privacy protection. We accomplish this by creating multiple identities for the same user and only allowing cross reference when necessary institutional regulations are accompanied with the data access.  
%
%
%We disguise the original user by multiple pseudo-ids that are ultimately cross reference-able to the original id.  Figure~\ref{fig:useraccess} shows different tables in our schema. The first table called \textit{user information table} contains the demographic information about the user, that is his age, country, his most used IP address and an id called as ``user global id". This table is never shared with anyone and the current recommendation is that it usually resides at a secure location with encryption.
\begin{figure}[ht!]
\centering
\includegraphics [width=0.8\textwidth]{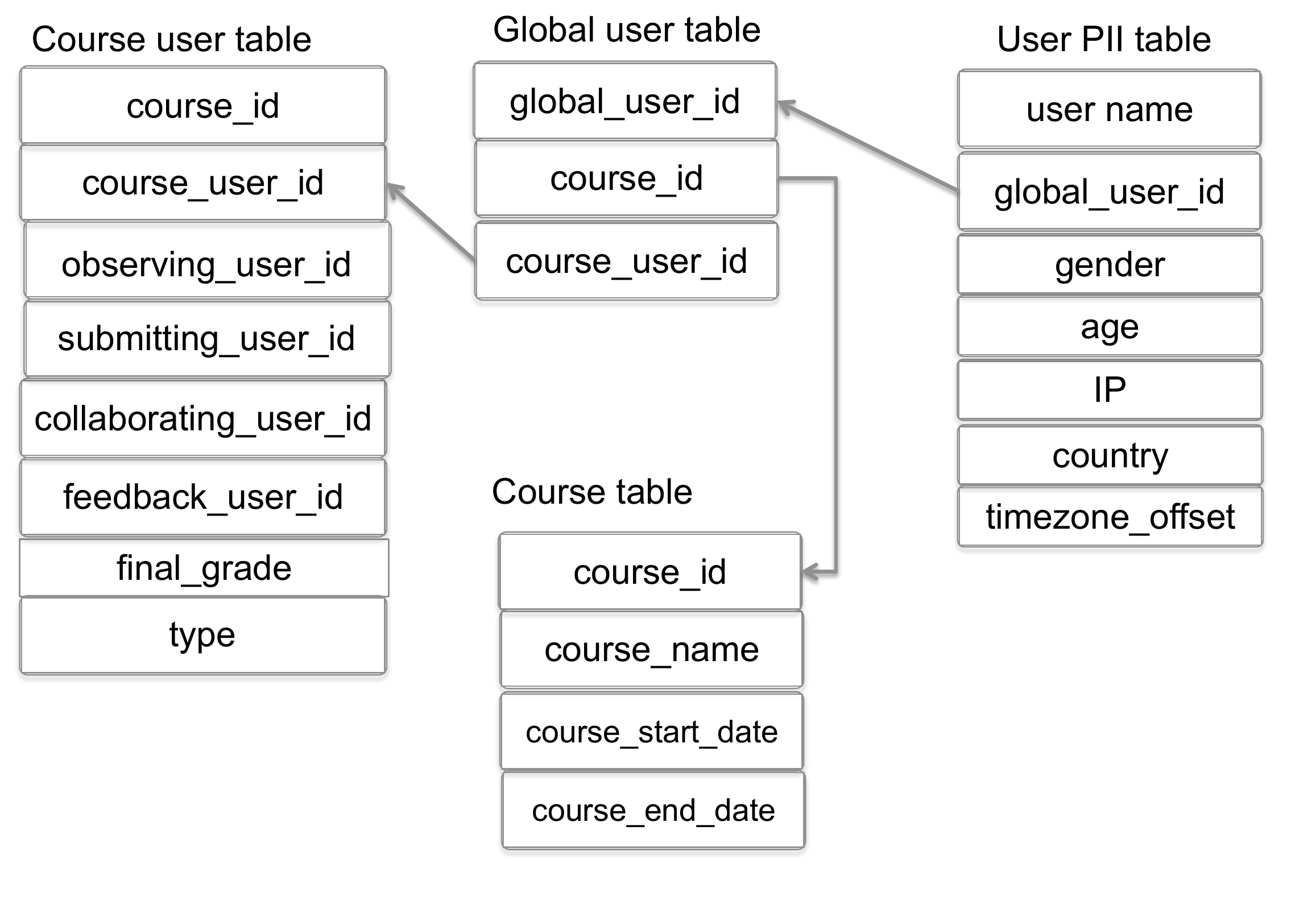}
\caption[User Information Tables] {Multiple layers of tables linking anonymized \id{s} allow access control over whether a user can be linked between courses or within a course across its different modes.}
\label{fig:useraccess}
\end{figure}

%The second table in this schema maps the global id to a unique course specific user id. For every course the user has taken under the MOOC provider the \textit{user\_id\_course} is a unique number that identifies this user with in that course. The course are identified by the $course\_id$ and its name, start and end dates are stored in the course information table. 
%
%The third table is the user table for the course. This table, in addition to the information pertaining to the type of the user and his final grade, maps the user's id within the course to his id within each of the other tables in the schema. For example, for the observed events table, the user is given an identifier which represents him uniquely within the scope of that table. This table allows to link the information pertaining to the same user across the tables.  

\begin{figure}[ht!]
\centering
\includegraphics [width=0.8\textwidth]{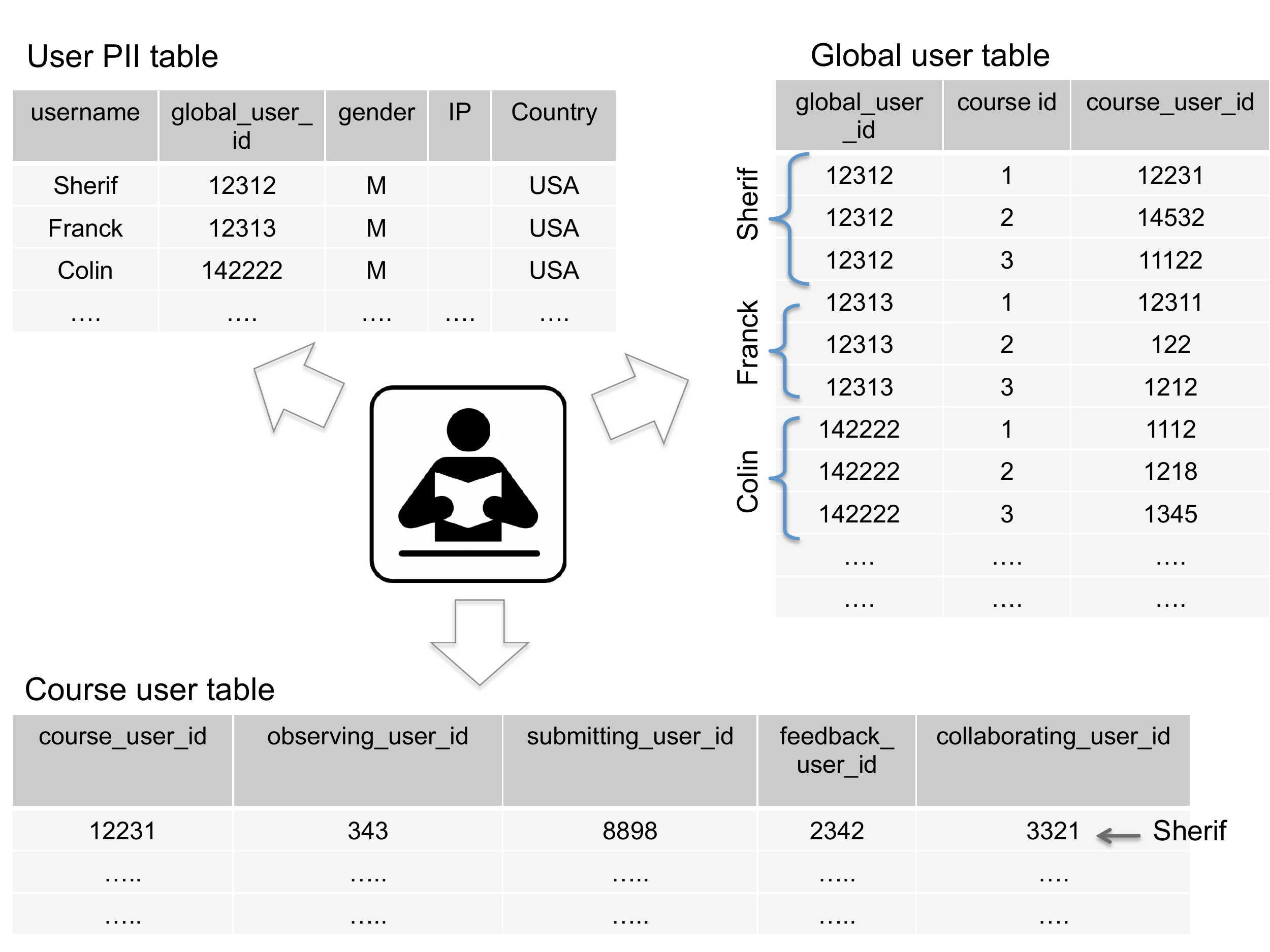}
\caption[User ID Cross Referencing] {User information tables and how they anonymize users in multiple layers -- across courses or within a course.}
\label{fig:useraccess}
\end{figure}

%\subsection{Privacy protecting access levels} 
The advantage of having multiple localized anonymized identifiers for users is that it supports a flexible access policy. Usually access to the highest level information, that has maximum possibility of user identifiability, is well guarded and there are restrictions to where the data can reside, who can use it and how the data will be stored. In almost all situations, a table such as the \textit{user PII table} is never shared.  A duly organized schema with multiple localized identifiers allows the data tables to be partitioned in multiple ways to grant multiple levels of access.

It is important to note however, even after masking the user with multiple anonymizing identities, there is some possibility the identity of a user will be identifiable because of the forum posts which are publicly available or other internet material beyond access control. For example, when a researcher is granted  access to the \textit{course user table} which cross references the user across multiple tables within the course, if there is a post on the forum (in the collaborating mode tables) that identifies a user, this post information could be combined with data in the \textit{course user table} to fully identify the specific actions of the user. This example implies that the subset of users who identify themselves on the forum are vulnerable to identification.  This may not be a large set of users but even one user's exposure of personal data is undesirable.  Hence an important consideration, even after limiting access to the \textit{user PII table},  is whether or not to give access to the collaborations table concurrently with other mode tables and user tables. Depending on access to the collaborations table, the schema then yields 7 different types of the access to the data which themselves depend on other restrictions to tables. Figure~\ref{fig:accesstypes} shows 6 different access types. They can be subdivided first on whether the data partition has the collaborating mode tables: 
 \begin{figure}[htb!]
\centering
\includegraphics [width=0.8\textwidth]{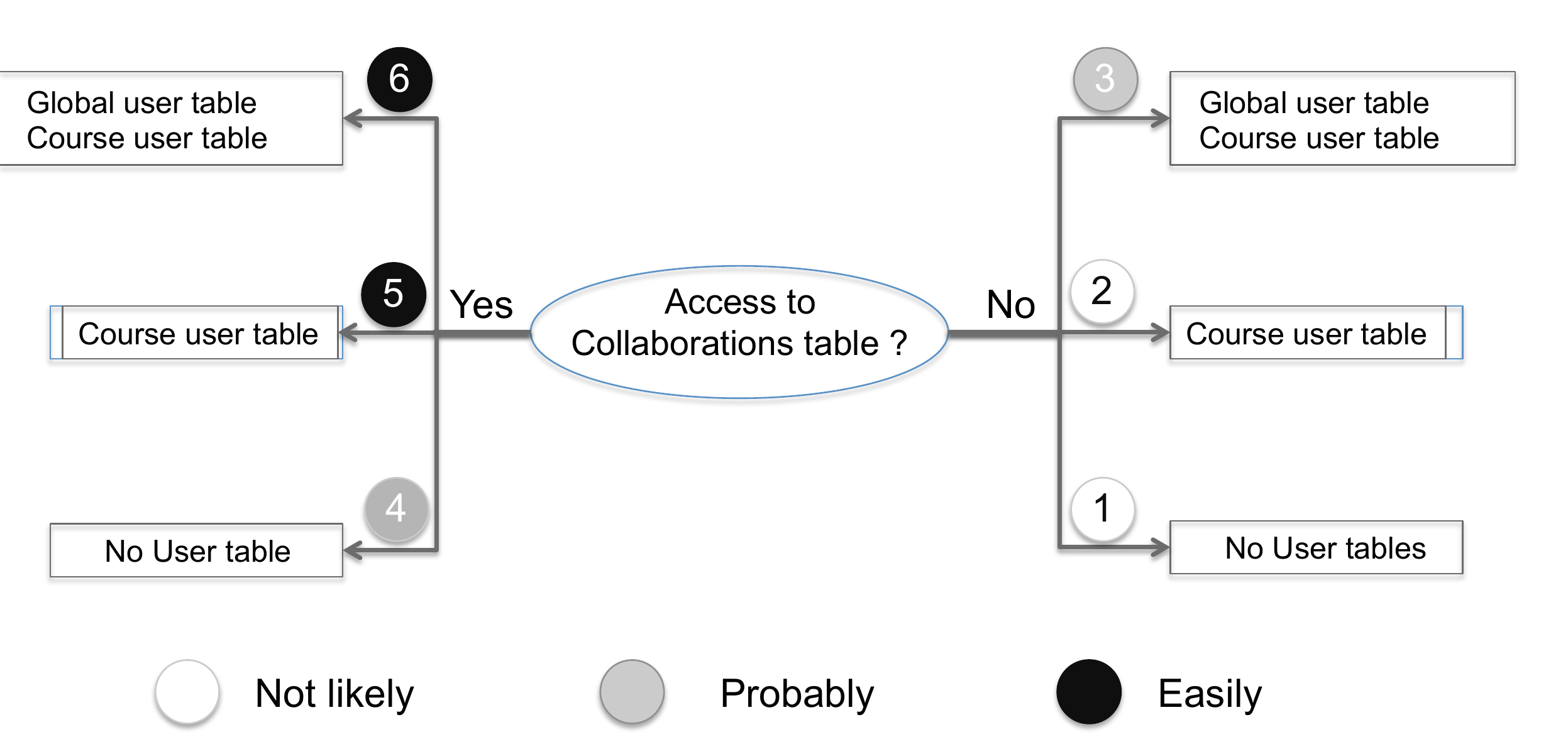}
\caption[Data Partitions and Their De-Identification Risk] {Data Partitions and Their De-Identification Risk.  Each partition equates with a level of access.  The different types of access are marked to show the degree to which it is likely that a user can re-identified. The most risky access allows access to the \textit{global user table} and the \textit{course user table}. }
\label{fig:accesstypes}
\end{figure}

\begin{enumerate}
\item {Without the collaborating mode tables:}
\begin{description}
\item \textbf{Multi course access}: At this level, the table that maps the \textit{global\_user\_id} to \textit{course\_user\_id},\textit{ global user table}  is shared along with the every course's \textit{course user table}. This allows researchers to track the user across multiple courses. Even though the \textit{user PII table} and collaborations mode tables are not provided across multiple courses, in some cases, with extremely hard work, a user can be identified. Such breaches usually happen when users themselves share a part of their information which is linked to some unique data item. 
\begin{exmp}
If a user shares which courses s/he has taken publicly and possibly some of their attributes, for example grade, with access to all the mode tables except collaboration  and with the ability to cross reference a user's information across courses, one can narrow down the number of users that map that profile and also likely identify a subset of students or, in very rare cases, a single user.
\end{exmp}
\item \textbf{Single course access}: At this level, the \textit{course user table} is shared, but the table that maps the user id at the course level to the global user id, \textit{global user table} is not shared. This does not allow researchers to track through multiple courses, but allows linking the information that pertains to the same user in multiple tables. Without the access to the collaborations table, it is highly unlikely that one can identify the user unless a user identifies several  rare or specific pieces of information about his behavior in the course. 

\item \textbf{Table level access}: At this level, the user table for the course is not shared. This leaves almost no possible way that the researchers can  link the information across multiple tables to the same user, thus achieving highest possible privacy protection in the current framework. We recommend that data at this level be guarded with minimal restrictions. 
\end{description}

\item{With the collaborating mode tables}:
\begin{description}
\item \textbf{Multi course access}: At this level, the table that maps the $user\_global\_id$ to $user\_course\_id$ is shared along with the entire user table for the course. This allows researchers to track the user across multiple courses. Though the username and other identifiable information is not made available, due to availability of the collaborations table , this allows identifiability of at least (if not all) the subset of the users who post on the forums. 

\item \textbf{Single course access}: At this level, the user table for a course is shared, but the table that maps the user id at the course level to the global user id is not shared. This does not allow researchers to track the user through multiple courses, but allows linking the information that pertains to the same user in multiple tables. Again due to the availability of collaborations table, a subset of users can be identified based on their forum posts.  

\item \textbf{Mode Table access}: At this level, the user table for the course is not shared. However the collaborations table is shared across the tables of different modes. The subset of users that post on the forums can be identified by using the time stamps that correspond to their posting and then identifying the corresponding event in the observed events table. This is because in our schema an access to the forums is also recorded as an event in the observed events table. 
\end{description}

\end{enumerate}

\moocDB will need to be flexible and readily  updatable to cope with the changing landscape around data control and privacy. Under FERPA and other regulations, some user data is considered ``education records'' and has to be protected to preserve privacy. Yet other data is confidential.
In general, a frequently used strategy to protect confidential and/or education records is to isolate, encrypt and restrict access to them under some transparent, formal policy. When access is granted, the \pii remains under restrictions but is under the control of the recipient. The recipient can consult the data and use it for research but is restricted from transferring it to any other party, from trying to de-identifying it, from disclosing anything about it which could lead to de-identification.

The issue with MOOC data currently, as it regards privacy protection, is that some of the data  could allow a user to be (de)identified but are not directly confidential or education records. One portion of this data is information solicited by the MOOC when the user signs up or that may be self-reported when a user fills out a demographic questionnaire. The second portion is the forum data. The forums are external to the platform and, while users are warned about doing it, allow users to reveal their identity. From a forum revelation, using timestamp correlation, it is, to some extent, possible to backtrack to the other data and identify a student in terms of behavioral path, i.e. the time they spent on resources, their answers to assessments, grades, etc.  The forum data which, in MOOCdb, is held in the collaborating mode tables \textbf{is also publicly available}.  This adds complexity to access considerations on MOOC data.  Given non \pii may allow a student to be identified within or external to the data, one means of mitigating this risk is to aggregate and/or ``noise'' the data before it is passed to the requester, to prevent re-identification. Effectively the data is modeled with specific noise integrated and surrogate data, from the model, can be provided. Some noise injection mechanisms are also accompanied by certificates of risk of re-identification. Other anonymizations of the data may be possible. For example, the timestamps could be randomized in a way that preserves ordering to some extent while offering anonymization.   We do not further cover these options in any more details  in our current report but certainly envision such approaches can help increase the protection of privacy while giving access to the data.

%\subsection{Minimal de-identified subset}
%Giving researchers table level access in which each user has a different identifier (id) in each of the tables is a promising way to provide the minimal subset of the data that is de-identified. It addresses the most important concerns where availability of linkage between the user id in the collaborations table and other tables allows identification of the user. This is because text entered in the collaborations table is publicly available because forums are publicly visible and users identity is revealed in the forums. However, since we record the users access of a forums page as an observed event as well,  one can argue that time stamps can be used to decipher the correlation between events in the observed mode table and the collaborations mode table. Hence we propose to be completely one should remove all the observed events that correspond to forum posts. Thus this set of data becomes the minimal de-identified subset. 

%\input{UserTables}
\vspace{-3mm}
\section{The edX 6.002x case study}\label{sect:useCase}
\vspace{-3mm}
edX offered its first course \textit{6.002x: Circuits and Electronics} in the Fall of 2012. 6.002x had 154,763 registrants. Of these, 69,221 people looked at the first problem set, and 26,349 earned at least one point on it. 13,569 people looked at the midterm while it was still open, 10,547 people got at least one point on the midterm, and 9,318 people got a passing score on the midterm. 10,262 people looked at the final exam while it was still open, 8,240 people got at least one point on the final exam, and 5,800 people got a passing score on the final exam. Finally, after completing 14 weeks of study, 7,157 people earned the first certificate awarded by MITx, showing that they successfully completed 6.002x.

\begin{figure}[ht]
\centering
\includegraphics [width=0.8\textwidth]{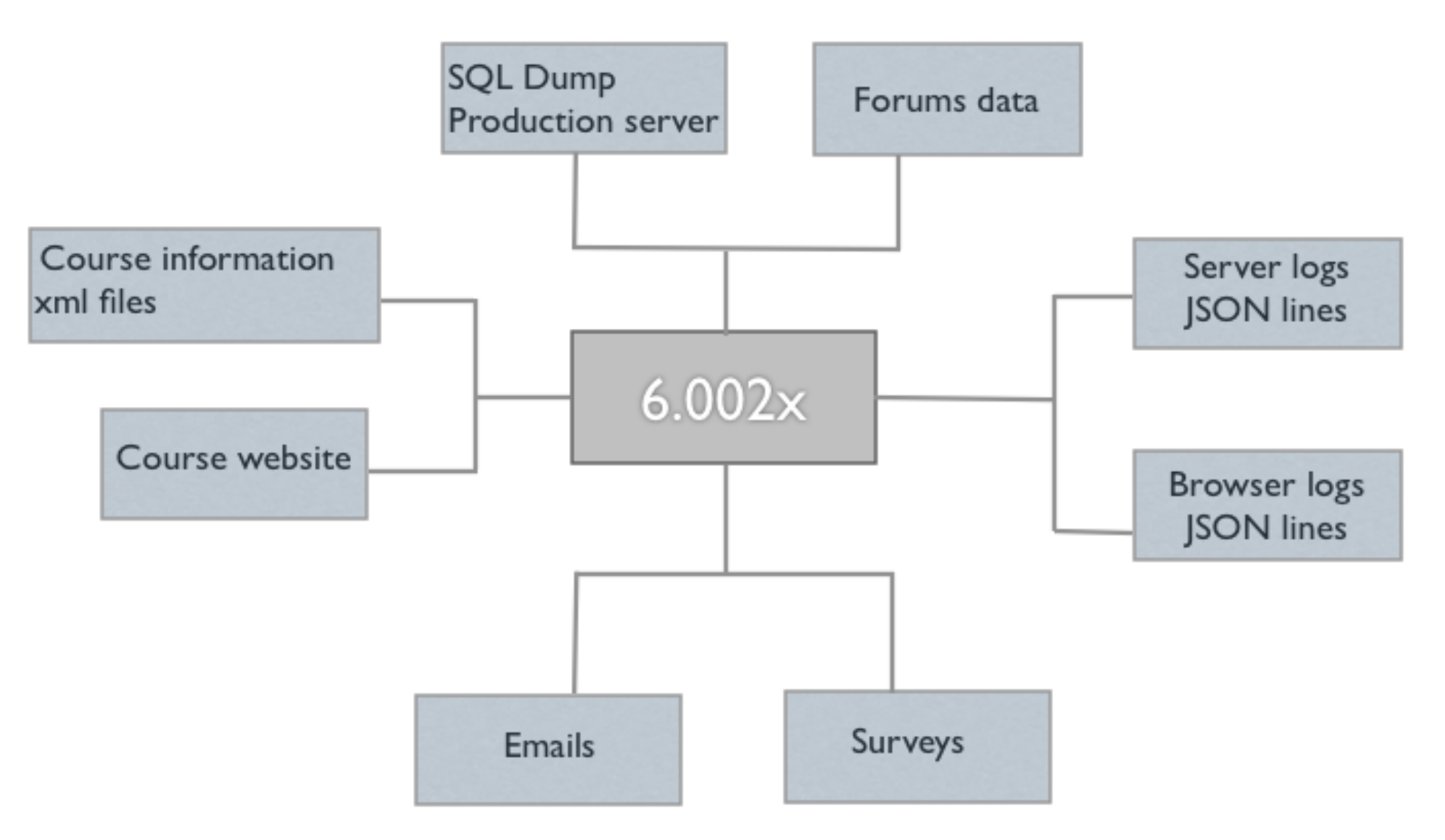}
\caption[Different data sources] {6.002x Data}
\label{fig:6.002xdata}
\end{figure}

The data corresponding to the student engagement was stored in different formats and was provided to us as shown in Figure~\ref{fig:6.002xdata} \footnote{We note this was the first offering of MITx and since then edX has update the storage formats}. The data formats comprised of XML files, JSON logs, sql database used in the production server, the course \textit{website}, \textit{forums} data, \textit{surveys} collected at the end of the course and \textit{emails}. To demonstrate the inefficiencies involved with this sort of heterogeneous data store, let us consider an example research question, ``How does amount of time spent on the videos during a certain week correlate to performance on the homework?". To evaluate this correlation one has to assemble a two dimensional array: The first column contains the amount of time spent on the video for every student for every week  and the second column contains the number of correct submissions for that student for the homework due that week. To get these two vectors, here are the steps one needs to take:

\begin{enumerate}
\item ``What is the time period between two consecutive home works?": This information is available by examining the home work release dates from website and the course xml files. 

\item ``Identify what are the video urls/uris correspond to these home works?" Between those dates, we next have to identify the urls/uris that correspond to videos and were available for the students. This information is available in the course xml files. 

\item ``Calculate the time spent on these urls by the user? " : We then calculate the time spent by the user on these urls. This can be calculated by processing the browser logs which capture the event logs per user as JSON lines. The script that does this calculation can be time consuming depending upon the number of entries in the tracking logs file. 

\item ``What answers did the user submit during this period?" This information per user can be attained from the events from the server logs. 

\item ``How many answers were actually correct?" For this we have to process the server logs to extract the answers the student submitted and then check whether they are right or not by querying the sql data base provided which records the correct answers per homework. 

\end{enumerate}
Once all these steps are finished and the two dimensional array is assembled, evaluating the correlation takes less than a minute by applying the \textit{core} function in MATLAB. This process is simply not scalable, it is fragile and computationally expensive. 

These original data pertaining to the observing mode was stored in files and when we transcribed in the database with fields corresponding to the names in the ``\textit{name-value}"  it was about the size of around 70 GB. We imported the data into a database with the schema we described in the previous subsections. The import scripts we had to build fell into two main categories:

\begin{itemize}
	\item reference generators, which build tables listing every user, resource and problem that were mentioned in the original data.
	\item table populators, which populate different tables by finding the right information and converting it if needed.
\end{itemize}

The sizes and the format of the resulting tables is as follows: submissions: 341 MB (6,313,050 rows); events: 6,120 MB (132,286,335 rows); problems: 0.08 MB; resources: 0.4 MB; resource types: 0.001 MB; users: 3MB. We therefore reduced the original data size by a factor of 10 while keeping most of the information. This allows us to retrieve easily and quickly information on the students' activities. For example, if we need to know what is the average number of pages in the book a student read, it would be around 10 times faster. Also, the relative small size of the tables in this format allows us to do all the work in memory on any relatively recent desktop computer. 

%\noindent \textbf{Why database?}: One of the questions we would like to answer is why organize the data in the form of the database even though it proves to be a barrier to scientists. The answer to this question is simply the scale at which things could be done. First the database schema allows to store the data in a concise format reducing the size. Second the analytics frameworks now available with databases allow data management and data extracting at lightning speeds. Most of the core technology behind databases is written in C. If one chooses to save the data in files and link them through scripts and progresses towards building a framework that could address many extraction procedures, it often results in building something that looks like a database albeit fragile and hard to maintain. 

\section{MOOCdb}\label{sect:tools}
\begin{figure}[ht]
\centering
\includegraphics [width=1\textwidth]{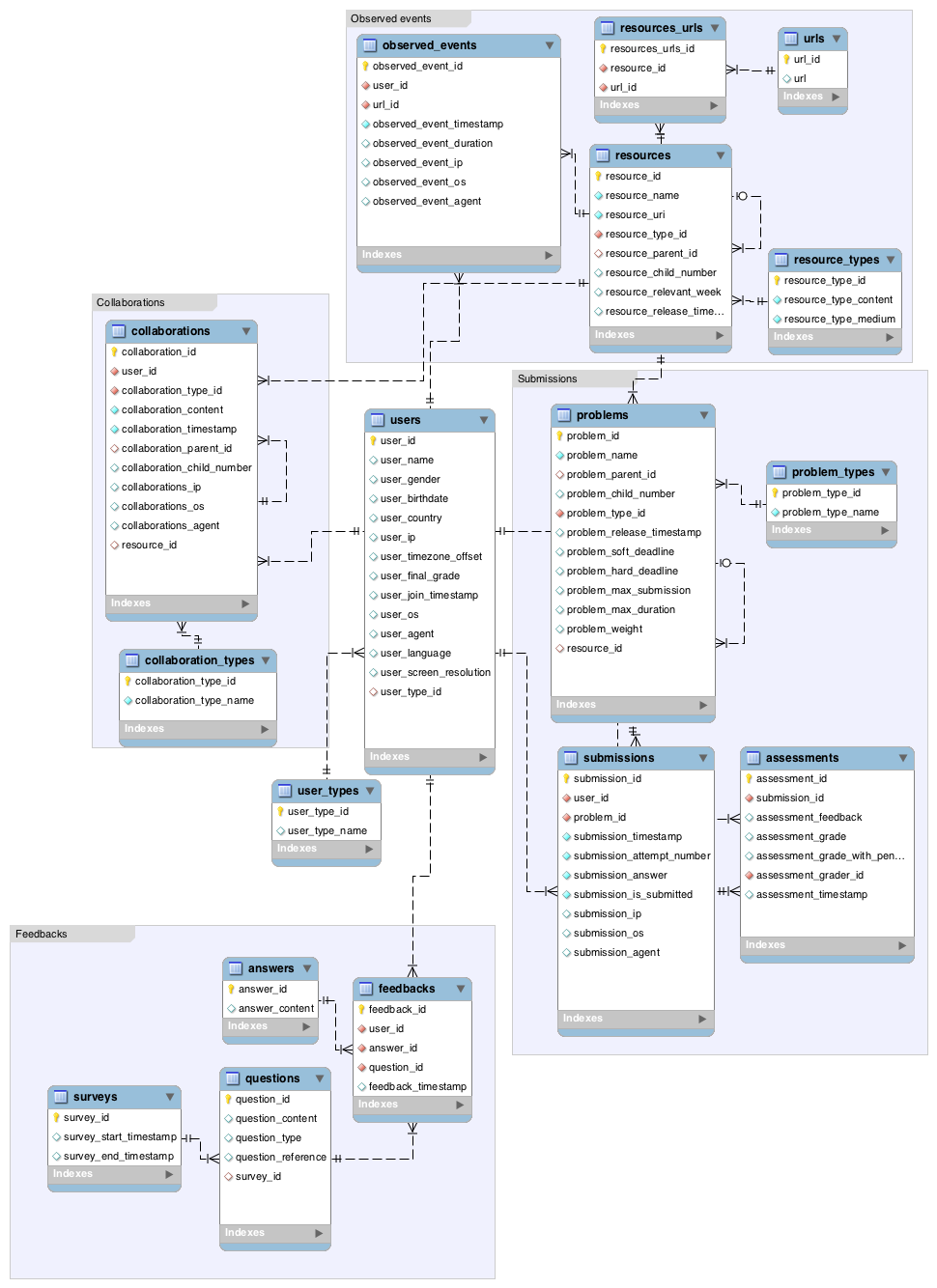}
\caption{The full database schema}
\label{fig:FullSchema}
\end{figure}

We present the schema in Figure~\ref{fig:FullSchema}. We surveyed a number of courses on both Coursera and edX and our schema can be populated with data from these multiple platforms. After organizing and storing this data concisely we built frameworks and tools that when put together form MOOCdb. The first is an analytics framework whose output can be plugged into different visualization techniques. The second provides scripts to create data exports for some specific research studies. The third framework develops tools that provide access to the DB from some well known statistical analysis packages.  

\begin{description}
\item \textbf{MOOC En Images}
MOOC En Images is an analytics framework that uses the schema. In this an analyst specifies a \textit{statistic} and defines the data over which this \textit{statistic} is calculated. The \textit{data} is defined by defining cuts along the three axes: \textit{time}, \textit{student cohort} and \textit{space} shown in Figure~\ref{fig:cuts} \footnote{A fourth axes currently considered is a \textit{resource} module}. An example visualization of an aggregate \textit{statistic} is shown in Figure~\ref{fig:example}. In the example, the aggregate statistic is \textit{the average number of home work submissions}. This statistic is then calculated for students who got the certificate from each country seperately. \footnote{This generalizable framework is currently under development and in its beta phase. We intend to release the framework}

\begin{figure}[ht!]
\centering
\includegraphics [width=0.65\textwidth]{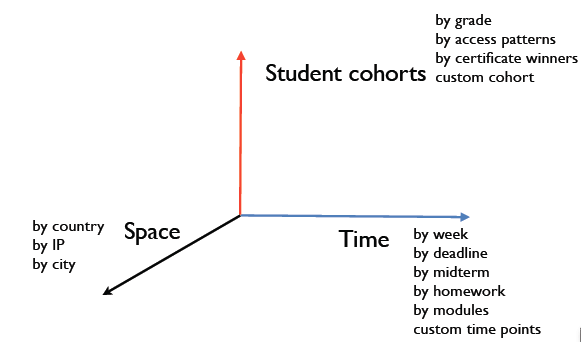}
\caption[Different cuts through the data] {Through the MOOC En Images analytics framework an analyst can cut the data along different axes. An analyst can define the a student cohort, a time period and the country/location if needed. The framework then is able to extract the data and evaluate the statistic the analyst is interested in and present it visually.}
\label{fig:cuts}
\end{figure}

\begin{figure}[ht!]
\centering
\includegraphics [width=0.65\textwidth]{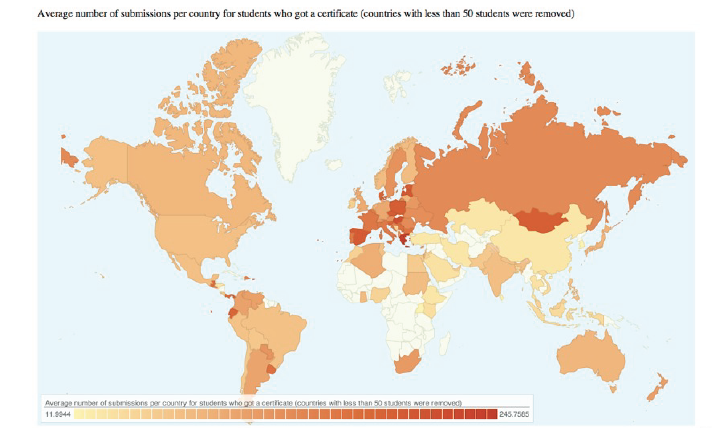}
\caption[An example visualization] {An example of visualization generated from data stored in the MOOCdb. The data is from the first offering of the MITx course: 6.002x. Here we show the \textit{average} number of submissions made by the students who got certificate calculated country-by-country. We see that Mongolia had a very high average.}
\label{fig:example}
\end{figure}

More detailed examples of visualizations of data created from the data are presented in \cite{MOOCEnImages}. To enable interesting attractive visualizations we are building tools that integrate visualization frameworks like \textit{flotr2},\textit{flot}, \textit{Dojo}, \textit{Google charts} and \textit{d3js}.

\item \textbf{Access tools}
\begin{figure}[ht!]
\centering
\includegraphics [width=1\textwidth]{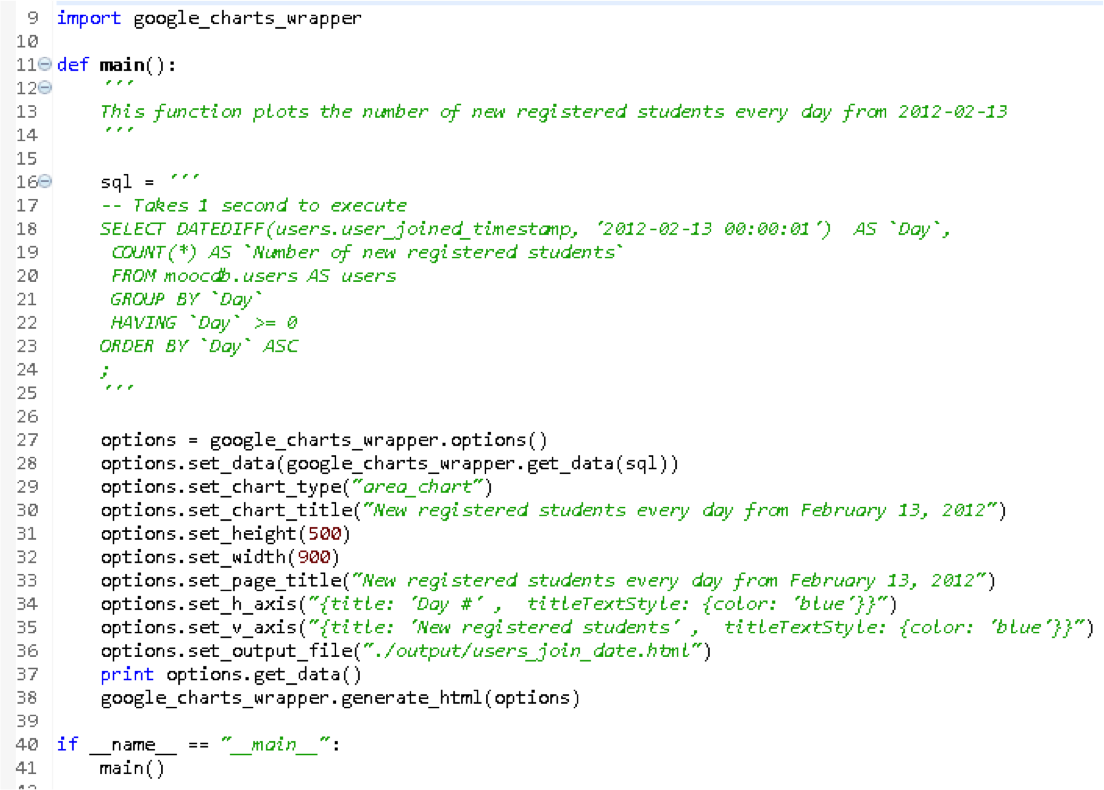}
\caption[Python code snippet] {Python code snippet. This is an example showing how the \textit{sql} queries can be called from Python.}
\label{fig:PythonCodeSnippet}
\end{figure}

\begin{figure}[ht!]
\centering
\includegraphics [width=1\textwidth]{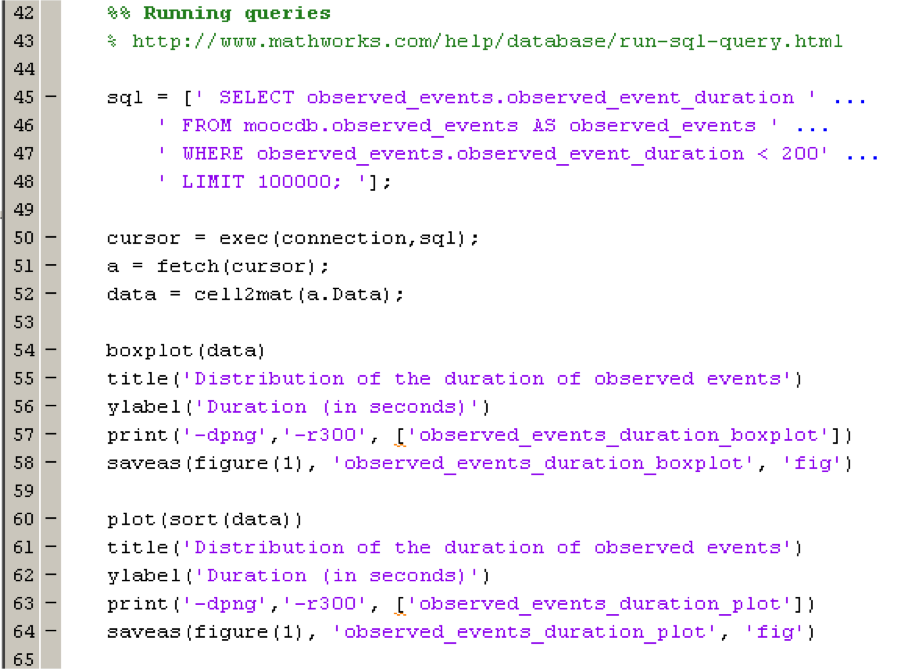}
\caption[MATLAB code snippet] {MATLAB code snippet. This is an example of how a the \textit{sql} queries written for MOOCdb can be called from MATLAB. }
\label{fig:MatlabCodeSnippet}
\end{figure}

To enable scientists to query the data easily without the need to know the database querying language, we built plugins for MATLAB, R, and Python. This allows the users to download the \textit{sql} queries from the public repository and use them in the language of their choice. This has two advantages. First the plugin allows users to easily access the data and second it allows processing the data in memory without the need to use the disk.

\item \textbf{Data exports}
Finally, for some standard research studies, researchers expect data in a certain format and stored in csv. Since the schema is standardized the scripts that will export the data for different studies are generated and shared. For our research purposes we are developing scripts that will extract the data for Bayesian knowledge tracing and item response theory. 

\end{description}

%Using some our new schema and some basic graphics packages, we 

%For more details about the analytics we performed as well as the entire database schema we refer the reader to \cite{dernoncourt} \footnote{ For the full MOOCdb database schema, see \url{http://bit.ly/MOOCdb} }

\vspace{-3mm}
\section{Conclusions and future work}\label{sect:conclusions}

In this paper, we proposed a standardized data schema and believe that this would be a powerful enabler for ours and others researchers involved in MOOC data science research. Currently, we after building databases based on this schema  for different courses and the accompanying analytic frameworks. We intend to release them in the near future. We believe it is timely to envision an open data schema for MOOC data science research. The future directions for this research involve: 

\noindent - Research into better efficient database technology and stores. \\
- Research into GPU based query execution, in-memory analytics.\\
- Research into how to store multiple courses. \\
- Meta tagging, provenance. \\
- Possibility of tracking student through multiple courses. \\

 Finally, we propose that as a community we should come up with a shared standard set of features that could be extracted across courses and across platforms. The schema facilities sharing and re-use of scripts. We call this the "feature foundry". In the short term we propose that this list is an open, living handbook available in a shared mode to allow addition and modification. It can be implemented as a google doc modified by the MOOC community. At the moocshop we would like to start synthesizing a more comprehensive set of features and developing the handbook. Feature engineering is a complex, human intuition driven endeavor and building this handbook and evolving this over years will be particularly helpful.

\bibliographystyle{plain}
\bibliography{moocPapers}

\end{document}